\documentclass[Journal]{IEEEtran}
\IEEEoverridecommandlockouts
\usepackage{graphicx, tikz, cite}
\usepackage{amsmath}
\usepackage{makeidx}
\usepackage{amsthm}
\usepackage{amssymb}
\usepackage{epstopdf}
\usepackage{multirow}
\usepackage{array}
\usepackage{color}
\usepackage{hyperref}
\usepackage{graphicx, psfrag, minibox, mathrsfs, bigints, stfloats, color, upgreek, gensymb, textcomp, euscript, calligra, xcolor, pict2e}
\date{}
\usepackage{diagbox}
\usepackage{slashbox}
\usepackage{algorithmicx}
\usepackage{algpseudocode}
\usepackage[ruled]{algorithm2e}
\usepackage{setspace}
\usepackage{float}
\usepackage[font=scriptsize]{caption}
\usepackage{subcaption}
\usepackage{enumerate}
\usepackage{eso-pic}
\definecolor{red}{rgb}{0.8,0.1,0.1}
\definecolor{blue}{rgb}{0,0,1}

\begin{document}
\title{Adaptive Transmission for Distributed Detection in Energy Harvesting Wireless Sensor Networks}
\author{\IEEEauthorblockN{Ghazaleh Ardeshiri, Azadeh Vosoughi~\IEEEmembership{Senior Member,~IEEE}} \thanks{Part of this work was presented in GLOBECOM 2019.}}
\maketitle
\begin{abstract}
We consider a wireless sensor network, consisting of $N$ heterogeneous sensors and a fusion center (FC), tasked with detecting a known signal in uncorrelated Gaussian noises. Each sensor can harvest randomly arriving energy and store it in a finite-size battery. Sensors communicate directly with the FC over orthogonal fading channels. Each sensor adapts its transmit symbol power, such that the larger its stored
energy and its quantized channel gain are, the higher its transmit symbol power is. To strike a balance between energy harvesting and energy consumption, we formulate two constrained optimization problems $(\mathcal{P}_1)$ and $(\mathcal{P}_2)$, where in both problems we seek  the jointly optimal local decision thresholds and channel gain quantization thresholds. While in $(\mathcal{P}_1)$ we maximize the $J$-divergence of the received signal densities at the FC, in $(\mathcal{P}_2)$ we minimize the average total transmit power, subject to certain constraints. We solve $(\mathcal{P}_1)$ and $(\mathcal{P}_2)$, assuming that the batteries reach their steady state. Our simulation results demonstrate the effectiveness of our optimization on enhancing the detection performance in $(\mathcal{P}_1)$, and lowering the average total transmit power in $(\mathcal{P}_2)$. They also reveal how the energy harvesting rate, the battery size, the sensor observation and communication channel parameters impact obtained solutions.   
\end{abstract}
\section{Introduction}
In a conventional wireless sensor network (WSN), sensors that are powered by non-rechargeable batteries and have limited  sensing, computation, and communication capabilities are used to sense and collect data for a wide range of applications \cite{Sudevalayam}. 
This energy constraint has inspired a rich body of research on developing signal processing and transmission strategies to achieve balance between network lifetime and performance (e.g., transmission rate, reliability, coverage) \cite{survey}, \cite{Mao}.
The technology of harnessing energy from the renewable resources of energy in ambient environment (e.g., solar, wind, and geothermal energy) has attracted attention of many researchers, as a promising solution to address the challenging energy constraint problem \cite{advances}. In particular, energy harvesting (EH)-powered sensors offer potential for transforming design and performance of wireless data communication systems and WSNs tasked with detection or estimation of a signal source. 
\par In practice, the energy arrival of ambient energy sources is intrinsically time-variant and often sporadic. This natural factor degrades the performance of the battery-free
EH-enabled wireless communication systems in which a harvest-then-transmit
strategy is adopted, i.e., users can only transmit when the energy harvested in one time slot is sufficient for data transmission. To flatten the randomness of the energy arrival, the harvested energy is stored in a battery, to balance the energy arrival and the energy consumption.
\par Power/energy management in EH-enabled communication systems with finite size batteries is necessary, in order to adapt the rate of energy consumption with the rate of energy harvesting. If the energy management policy is overly aggressive, such that the rate of energy consumption is greater than the rate of energy harvesting, the transmitter may stop functioning, due to energy outage. On the other hand, if the policy is overly conservative, the transmitter may fail to utilize the excess energy, due to energy overflow, and the data transmission would become limited in each energy allocation interval. In the following, we briefly discuss the most related literature on EH-powered wireless networks. 
\subsection{Literature Review}
The authors in \cite{Tarighati},\cite{khar1} studied design and performance of WSNs, consisting of EH-powered sensors, that are tasked with solving a binary hypothesis testing problem. 
Modeling the battery state as a two-state Markov chain and choosing Bhattacharya distance as the detection performance metric, the authors in \cite{Tarighati} investigated the optimal local decision thresholds at the sensors, such that the detection performance is optimized, subject to the causality constraint of the battery. 
Choosing error probability as the detection performance metric, 
the authors in \cite{khar1} studied ordered transmission schemes, and proposed schemes that can lead to a smaller average number of transmitting sensors, compared with the unordered  transmission scheme, without comprising the detection performance.
Considering an EH-powered node, that is deployed to monitor the change in its environment, the authors in \cite{khar2} formulated a quickest change detection problem, where the goal is to detect the time at which the underlying distribution of sensor observation changes. Taking into account the energy consumption for making observations, the authors in \cite{khar2} studied the optimal stopping time at the sensor, such that the worst case detection delay is minimized, subject to different average run length constraints.
\par The authors in \cite{nourian} addressed distributed estimation of a random signal source using multiple EH-powered sensors in a WSN. Aiming at minimizing the total distortion over a finite time horizon or a long term average distortion over an infinite time horizon, subject to energy harvesting constraints, the authors in \cite{nourian} explored optimal energy allocation policies at sensors. Considering a WSN, consisting of EH-enabled sensors, among which one sensor is selected to transmit to the fusion center (FC), the authors in \cite{khairnar} jointly optimized sensor selection and selected sensor's discrete rate, such that the average throughput over an infinite time horizon is maximized, subject to a target BER constraint.
The author in \cite{hong} considered a WSN that is tasked with estimation of a deterministic signal source, using multiple EH-powered nodes and a FC. Assuming that each node performs amplify-and-forward (AF) relaying toward the FC, and the FC applies maximum likelihood estimation, the author derived the optimal scaling factor at each node, such that the Fisher information of the unknown source is maximized. 
The authors in \cite{liu} proposed  a game-theoretic approach to model the distributed estimation problem in an energy harvesting WSN, where each sensor (as a player in the game) desires to maximize its own utility, given the complete information about other players' utilities and strategies.
\par Considering a cooperative data communication system, in which EH-powered relays volunteer to serve as AF relays, 
the authors in \cite{medepally} characterized the fading-averaged symbol error rate of the system for M-PSK constellation. 
 Considering a cooperative data communication system similar to that of \cite{medepally}, the authors in \cite{lioutage} approximated the system outage probability with one active relay, and showed that this probability is smaller, compared with that of simple direct source-destination transmission protocol. Considering a $K$-user multiple access channel with EH-enabled  transmitters, the authors in \cite{wang} investigated offline energy scheduling schemes over a finite number of time slots, with the goal of maximizing the sum-rate, subject to certain 
 constraints, and showed that the optimal scheme is obtained via an iterative dynamic water-filling algorithm.
\par Energy harvesting has also been considered in the context of cognitive radio systems, where the researchers have assumed that the secondary user transmitters (SU$_{tx}$) are capable of harvesting energy  \cite{ahmed}, \cite{park}, \cite{taherpour}, \cite{park2}, \cite{yazdani2020}. Aiming at maximizing the average throughput of SU$_{tx}$'s, the authors in \cite{ahmed,park} optimized the threshold for detecting the primary user (PU) signal,  considering the causality constraint of the battery and the collision probability constraint (which limits the probability of accessing the occupied spectrum) \cite{park}. 
The authors in \cite{taherpour} optimized the threshold for detecting the PU signal, such that the error probability of detecting the PU signal is minimized. Targeting at maximizing the average throughput of SU$_{tx}$s, the authors in \cite{park2} jointly optimized the threshold for detecting the PU signal and the duration of spectrum sensing. In \cite{yazdani2020}, the authors designed an energy management strategy that maximizes the achievable sum-rate of SUs, taking into account the combined effects  of spectrum sensing error and imperfect
channel state information (CSI), and subject to an interference power constraint and the causality constraint of the battery.
\subsection{Our Contribution}
We consider the distributed detection of a known signal in uncorrelated Gaussian noises (a binary hypothesis testing problem) using a WSN, consisting of $N$ EH-powered heterogeneous sensors and a FC.
Each sensor is equipped with a finite-size battery, to store the harvested energy from the ambient environment.
Sensors have access to their quantized channel gains, process locally their noisy observations, and communicate directly with the FC over orthogonal fading channels. In particular, sensor $n$ compares its test statistic against a local decision threshold $\theta_n$. If the test statistic exceeds $\theta_n$, sensor $n$ chooses its non-negative transmission symbol $\alpha_{n,t}$ according to the amount of available stored energy in its battery and its quantized channel gain. The FC applies the optimal Bayesian fusion rule to make the final decision about the presence or absence of the known signal. Our goal is to develop a power adaptation scheme for sensors  that strikes a balance between energy harvesting and energy consumption for data transmission, and optimizes the system performance. To achieve our goal we formulate two constrained optimization problems $(\mathcal{P}_1)$ and $(\mathcal{P}_2)$, where in both problems we seek the jointly optimal local decision thresholds $\theta_n, \forall n$ 
and channel gain quantization thresholds $\{\mu_{n,l}\}_{l=1}^L, \forall n$. 
\\ In the first problem $(\mathcal{P}_1)$, we optimize the detection performance corresponding to the decision at the FC, subject to an average transmit power per sensor constraint. In the second problem $(\mathcal{P}_2)$, we minimize
the average total transmit power, subject to the detection performance per sensor constraint.
we choose the $J$-divergence between the distributions of the detection statistics at the FC under two hypotheses, as the detection performance
metric \cite{vin}. Our choice is motivated by the facts that (i) it is a widely used metric for evaluating detection system performance \cite{vin}, \cite{Guo}, \cite{zahra}, since it provides a lower bound on the detection error probability. Hence, maximizing the $J$-divergence is equivalent to minimizing
the lower bound on the error probability; (ii) it is closely related to other types of detection performance metric, including  the asymptotic relative efficiency (ARE) \cite{Goodman}, Bhattacharya distance \cite{Tarighati}, and Kullback Leibler distance \cite{Ardeshiri}; (iii) it allows us to provide a more
tractable analysis.
 \\ To the best of our knowledge, our work is the first to study design and performance of a WSN with EH-powered heterogeneous sensors that consider {\it both} the battery state and the partial channel gain knowledge, to adapt transmission power, for optimizing the detection performance at the FC. None of the works in \cite{Tarighati}, \cite{khar1}, \cite{khar2} have considered partial CSI at sensors.
The importance of our study is evident by the works in \cite{ahmadi1}, \cite{ahmadi2}, \cite{ahmadi3} which demonstrate the significance of considering the effect of partial CSI at sensors on the detection performance at the FC. We note that sensors in  \cite{ahmadi1}, \cite{ahmadi2}, \cite{ahmadi3} have traditional stable power supplies.  One expects that partial CSI, combined with random energy arrival at each sensor, would impact the overall design and performance analysis problem in hand, and introduce new challenges that have not been addressed before. Our main contributions can be summarized as follow:
 \begin{itemize}
     \item Our system model encompasses the stochastic energy arrival model for harvesting energy, and the stochastic energy storage model for the finite-size battery. We model the randomly arriving energy units during a time slot as a Poisson process, and the dynamics of the battery as a finite state Markov chain.
     \item We propose a power adaptation scheme, which allows each sensor to first test the significance of its observation, and then to intelligently choose its transmission symbol, such that the larger its stored energy and its quantized channel gain are, the  higher its transmit power is. 
The optimization parameters $\theta_n, \{\mu_{n,l}\}_{l=1}^L, \forall n$, are embedded in the proposed power adaptation scheme, and  play key roles in balancing energy harvesting and energy consumption for channel-aware data transmission.
     \item Given our system model, we find the $J$-divergence and the average total transmit power, we formulate two novel constrained optimization problems $(\mathcal{P}_1)$ and $(\mathcal{P}_2)$, and we solve these problems assuming that the battery reaches its steady-state.
     \item We provide two approximations for the error probability corresponding to the optimal Bayesian fusion rule at the FC, relying on low signal-to-noise ratio (SNR) approximation for the communication channel noise, and Lindeberg Central Limit Theorem (CLT) for large $N$. 
 \end{itemize}
\par This work is different from our previous works in \cite{Ardeshiri}, \cite{Ard2}. In particular, in \cite{Ardeshiri} we approached the detection problem from Neyman-Pearson perspective. Assuming perfect CSI at sensors, we sought the optimal local decision thresholds such that the Kullback-Leibler distance between the distributions of the binary  detection statistics at the FC is maximized. In \cite{Ard2}, we adopted the Bayesian perspective and we explored the optimal transmit power map at sensors, such that $J$-divergence between the distributions of the binary detection statistics at the FC is maximized, subject to certain constraints. 
\subsection{Paper Organization}
The paper organization follows: Section \ref{sym} describes our system and observation models and 
introduces our two proposed constrained optimization problems $(\mathcal{P}_1)$ and $(\mathcal{P}_2)$. Section \ref{J_error} derives a closed-form expression for the total $J$-divergence and provides two approximations for the error probability corresponding to the optimal Bayesian fusion rule at the FC. Section \ref{cost_fun} derives the cost functions of $(\mathcal{P}_1)$ and $(\mathcal{P}_2)$ problems.  Section\ref{random_dep} discusses how our setup can be extended to the scenario where sensors are randomly deployed in the field, and hence sensors' locations are unknown.
 Section \ref{simulation} illustrates our numerical results. Section \ref{conclu} includes our concluding remarks. 
\section{System Model}\label{sym}
Our system model consists of $N$  spatially distributed sensors and a fusion center (FC) that is tasked with solving a binary hypothesis testing problem. Each sensor is capable of harvesting energy from the ambient environment and is equipped with a battery of finite size to store the harvested energy. Sensors process locally their observations and communicate directly with the FC over orthogonal fading channels. 
\subsection{Observation Model at Sensors}
\begin{figure}[!t]
\begin{subfigure}[t]{0.5\textwidth}
  \centering
  \includegraphics[scale=.33]{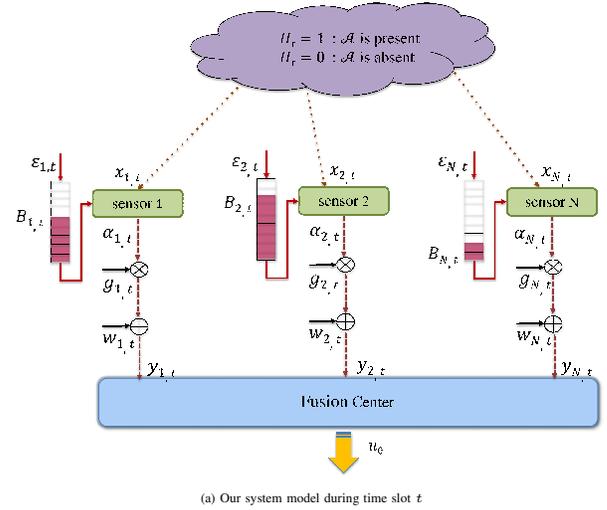}  
  \caption{Our system model during time slot $t$ }
 \label{}
\end{subfigure}
\begin{subfigure}[t]{0.5\textwidth}
  \centering
  \includegraphics[scale=.27]{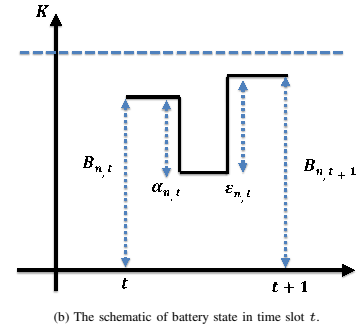} 
 \caption{The schematic of battery state in time slot $t$.}
  \label{}
\end{subfigure}
\caption{Our system model and the schematic of battery state in time slot $t$.}
\label{}
\end{figure}
To describe our signal processing blocks at sensors and the FC as well as energy harvesting model, we divide time horizon into slots of equal length $T_s$. Each time slot is indexed by an integer $t$ for $t=1,2,...,\infty$. We model the underlying binary hypothesis $H_t$ in time slot $t$ as a binary random variable $H_t \in \{0,1\}$ with a-priori probabilities $\Pi_0=\Pr(H_t=0)$ and $\Pi_1=\Pr(H_t=1)=1-\Pi_0$. We assume that the hypothesis $H_t$ varies over time slots in an independent and identically distributed (i.i.d.) manner. Let $x_{n,t}$ denote the local observation at sensor $n$ in time slot $t$. We assume that  sensors' observations given each hypothesis with conditional distribution $f(x_{n,t}|H_t=h_t)$ for $h_t \in \{0,1\}$ are independent across sensors. This model is relevant for WSNs that are tasked with detection of a known signal in uncorrelated Gaussian noises with the following signal model
\begin{align}\label{xk}
&H_{t}=1:~~ x_{n,t} ={\cal A}+v_{n,t},\nonumber\\
&H_{t}=0:~~ x_{n,t} = v_{n,t},~~
\text{for}~n=1,\dots,N
\end{align}
where Gaussian observation noises $v_{n,t} \! \sim \! {\cal N}(0,\sigma_{v_{n}}^2)$ are independent over time slots and across sensors. 
Given observation $x_{n,t}$ sensor $n$ finds local log-likelihood ratio (LLR)
\begin{equation}\label{lrt_sensor}
 \Gamma_n(x_{n,t})\triangleq\log\left( \frac{f(x_{n,t}|h_{t}=1)}{f(x_{n,t}|h_{t}=0)}\right ),   
\end{equation}
and uses its value to choose its non-negative transmission symbol $\alpha_{n,t}$ to be sent to the FC.
In particular, when LLR is below a local threshold $\theta_n$ (to be optimized), sensor $n$ does not transmit and let $\alpha_{n,t}=0$. When  LLR exceeds the local threshold $\theta_n$, sensor $n$ chooses $\alpha_{n,t}$ according to its battery state and the quality of sensor-FC communication channel (choice of $\alpha_{n,t}$ will be explained later).
\subsection{Battery State, Harvesting and Transmission Models
}
We assume sensors are equipped with identical batteries of finite size $K$ cells (units), where each cell corresponds to $b_u$ Jules of stored energy. Therefore, each battery is capable of storing at most $K b_u$ Jules of harvested energy. Let $B_{n,t} \in \{0,1,...,K\}$ denote the discrete random process indicating the battery state of sensor $n$, and $b_{n,t}$ represent the actual battery state (i.e., the actual amount of available energy units) at the beginning slot $t$. Note that $b_{n,t}=0$ and $b_{n,t}=K$ represent the empty battery and full battery levels, respectively.
Let ${\cal E}_{n,t}$ denote the randomly arriving energy units\footnote{Suppose each arriving energy unit measured in Jules is $b_u$ Jules.} during time slot $t$ at sensor $n$. We assume ${\cal E}_{n,t}$'s are i.i.d. over time slots and across sensors. We model ${\cal E}_{n,t}$ as a Poisson random variable with parameter $\rho$, and probability mass function (pmf) $p_e \triangleq\Pr({\cal E}_{n,t}=e)= e^{\rho}\rho^e/e!$ for $e=0,1,\dots,\infty$. 
Note that $\rho$ is the average number of arriving energy units during one time slot at each sensor. Let $e_{n,t}$ be the number of stored (harvested) energy units in the battery at sensor $n$ during time slot $t$. Note that the harvested energy $e_{n,t}$ cannot be used during slot $t$. 
Since the battery has a finite capacity of $K$ cells, $e_{n,t} \in \{0,1,...,K\}$. Also, $e_{n,t}$ are i.i.d. over time slots and across sensors. We  can find the pmf of $e_{n,t}$ in terms of the pmf of ${\cal E}_{n,t}$. Let $q_e\triangleq \Pr(e_{n,t}=e)$ for $e=0,1,\dots,K$. We have
\begin{equation}
    q_e=\begin{cases}
    p_e,~~~~~~~~~~~~~~\text{if}~ 0\leq e \leq K-1,\\
    \sum_{m=K}^\infty p_m,~~~~~\text{if}~ e=K.
    \end{cases}
\end{equation}
\par Let $g_{n,t}$ indicate the fading channel gain between sensor $n$ and the FC during time slot $t$. 
We assume block fading model and $g_{n,t}$'s  are  i.i.d.  over time slots and independent across sensors. We consider a coherent FC with the knowledge of all channel gains and assume that the FC quantizes the channel gain $g_{n,t}$ of sensor $n$ in time slot $t$. In particular, suppose the FC partitions the positive real line into $L$ disjoint intervals $\mathcal{I}_{n,1},..., \mathcal{I}_{n,L}$ using the quantization thresholds $\{ \mu_{n,l}\}_{ l=1}^L$, where $0\! = \mu_{n,0}\! < \mu_{n,1}\! < \dots \!< \mu_{n,L+1}\! = \infty$  (to be optimized). The feedback from the FC to sensor $n$ conveys the information to which interval $g_{n,t}$ belongs. 
We define the probability 
$\pi_{n,l}=\Pr(g_{n,t}\in \mathcal{I}_{n,l})$, which can be found based on the distribution of fading model in terms of the two quantization thresholds $\mu_{n,l}$ and $\mu_{n,l+1}$. For instance, for Rayleigh fading model $g_{n,t}^2$ has exponential distribution with the mean $\mathbb{E}\{g_{n,t}^2\}=\gamma_{g_n}$ and we have 
\begin{equation}\label{gaz}
\Pr\Big (g_{n,t}^2 \in (\mu_l^2, \mu_{l+1}^2)\Big )=\pi_{n,l}= e^{-\mu_{n,l}^2/\gamma_{g_n}}-e^{-\mu_{n,l+1}^2/\gamma_{g_n}}.
\end{equation}
In time slot $t$, if LLR exceeds the local threshold $\theta_n$, sensor $n$ chooses its non-negative transmission symbol $\alpha_{n,t}$ according to its battery state $b_{n,t}$ and the feedback information available on the channel gain $g_{n,t}$, such that symbol transmit power is higher when the channel gain is larger. Given $c_1, c_2, \dots, c_{L} \in [0,1]$ and $c_{l}> c_{l-1}$ for $l=1,...,L$, we adopt the following transmit power control strategy
\begin{equation}\label{alpha}
\alpha_{n,t}^2 =
\begin{cases}
0,&~~~\Gamma_n(x_{n,t})<\theta_n,\\
\lfloor c_1 b_{n,t}\rfloor,&~~~\Gamma_n(x_{n,t})\geq\theta_n,~ g_{n,t}\in\mathcal{I}_{n,1},\\
\vdots&~~~~~~~~~~~~~\vdots\\
\lfloor c_{L} b_{n,t}\rfloor ,&~~~\Gamma_n(x_{n,t})\geq\theta_n,~g_{n,t}\in\mathcal{I}_{n,L}.
\end{cases}
\end{equation}
where $\lfloor . \rfloor$ is the floor function. Assuming the consumed energy for sensing is negligible, the battery state at the beginning of slot $t+1$ depends on the battery state at the beginning of slot $t$, the harvested energy during slot $t$, and the transmission symbol $\alpha_{n,t}$ (see Fig. 2), i.e.,
\begin{equation}\label{b_n,t}
    B_{n,t+1} = \min\big\{[B_{n,t} + e_{n,t}-\alpha_{n,t}]^+,K \big \},
\end{equation}
where $[x]^+=\max \{0,x\}$. Considering the dynamic battery state model in \eqref{b_n,t} we note that, conditioned on $e_{n,t}$ and $\alpha_{n,t}$ the value of $B_{n,t+1}$ only depends on the value of $B_{n,t}$ (and not the battery states of time slots before $t$). Hence, the process $B_{n,t}$ can be modeled as a Markov chain. Let $\Phi_{n,t}$ be the probability vector of battery state in slot $t$
\begin{equation}\label{battery_state}
  \boldsymbol{\Phi}_{n,t}\triangleq\Big[\Pr(B_{n,t}=0),\dots,\Pr(B_{n,t}=K)\Big]^T,
\end{equation}
where the superscript $T$ indicates transposition. We note that $\Pr(B_{n,t}=k)$ in \eqref{battery_state} depends on $B_{n,t-1}$, $e_{n,t-1}$ and $\alpha_{n,t-1}$. Assuming that the Markov chain is time-homogeneous, we let $\Psi_n$ be the corresponding $(K+1) \times (K+1)$ transition probability matrix of this chain with its $(i,j)$-th entry 
$\psi_{i,j}\triangleq\Pr(B_{n,t}=j|B_{n,t-1}=i)$ for $i,j = 0,\dots, K$. 
\begin{figure}[!t]
	\centering
	\includegraphics[scale=.22]{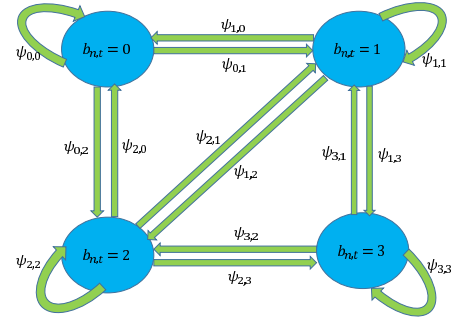}
	\caption{The schematic representation of 4-state Markov chain}
	\label{fig_battery}
\end{figure}
 Defining the indicator function $I_{i\rightarrow j}(e_{n,t},\alpha_{n,t})$ as the following
\begin{equation}\label{indicator}
    I_{i\rightarrow j}(e_{n,t},\alpha_{n,t})=
    \begin{cases}
    1,&\text{if}~ j\!=\!\min \big\{[i+e_{n,t}-\alpha_{n,t}]^+,K\big\},\\
    0,&\text{o.w.}
    \end{cases}
\end{equation}
we can express $\psi_{i,j}$ as below

%
%
\begin{align}
    \psi_{i,j}\!=\! \widehat{\Pi}_{n,1}\sum_{e=0}^K \sum_{l=1}^{L}\pi_{n,l}q_e I_{i\rightarrow j}(e_{n,t},\lfloor c_l i\rfloor)\nonumber
\end{align}    
\begin{align}\label{pij}
    & \!\!\!\!\!\!\!\!\!\!\!\!\!+\!\widehat{\Pi}_{n,0}\!\sum_{e=0}^K q_e I_{i\rightarrow j}(e_{n,t},0).
\end{align}
%
The symbols $\widehat{\Pi}_{n,0}$ and $\widehat{\Pi}_{n,1}$ in \eqref{pij} refer to the probabilities of events $\alpha_{n,t} = 0$ and $\alpha_{n,t} \neq 0$, respectively. In particular, we have
\begin{align}
    &\widehat{\Pi}_{n,0} = \Pr(\alpha_{n,t}\! =\! 0) = \Pi_0(1\!-\!P_{\text{f}_n}) + \Pi_1(1\!-\! P_{\text{d}_n}),\nonumber\\
    &\widehat{\Pi}_{n,1}= \Pr(\alpha_{n,t}\! \neq \! 0) = \Pi_0 P_{\text{f}_n} + \Pi_1 P_{\text{d}_n},
\end{align}
where the probabilities $P_{\text{f}_n}$ and $P_{\text{d}_n}$ can be determined using our signal model in \eqref{xk}
\vspace{-2mm}
\begin{align}\label{pd_pf1}
 &P_{\text{f}_n} \!= \! \Pr(\alpha_{n,t}\!\neq\!0 |h_t=0)\!=\!Q\Big(\frac{\theta_n+{\mathcal{A}^2}/{2\sigma^2_{v_n}}}{\sqrt{\mathcal{A}^2/{\sigma^2_{v_n}}}}\Big), \nonumber\\
 &P_{\text{d}_n} \!= \!\Pr(\alpha_{n,t}\!\neq \!0 |h_t=1)\!=\!Q\Big(\frac{\theta_n-{\mathcal{A}^2}/{2\sigma^2_{v_n}}}{\sqrt{\mathcal{A}^2/{\sigma^2_{v_n}}}}\Big). 
\end{align}
Since the Markov chain characterized by the transition probability matrix $\Psi_n$ is irreducible and aperiodic, there exists a unique steady state  distribution, regardless of the initial state \cite{shortle}. Let $\Phi_n=[\phi_{n,0}, \phi_{n,1}, ...,\phi_{n,K}]^T$ be the unique steady state probability vector with the entries $\phi_{n,k}=\lim _{t\rightarrow \infty} \Pr(B_{n,t}=k)$. Note that this vector satisfies the following eigenvalue equation 
\begin{equation}\label{inf}
 \boldsymbol{\Phi}_{n}=\Psi_n \boldsymbol{\Phi}_{n}.
\end{equation}
In particular, we let $\boldsymbol{\Phi}_{n}$ be the normalized eigenvector of $\Psi_n$ corresponding to the unit eigenvalue, such that the sum of its entries is one \cite{Tarighati}. The closed-form expression for $ \boldsymbol{\Phi}_{n}$ is 
\begin{equation}
 \boldsymbol{\Phi}_{n} = (\Psi_n-\textbf{I}-\textbf{B})^{-1}\textbf{1},
\end{equation}
where $\textbf{B}$ is an all-ones matrix, $\textbf{I}$ is the identity matrix, and \textbf{1} is an all-ones column vector. 
\\For clarity of the presentation and to illustrate our transmission model in \eqref{alpha}, we consider the following simple example consisting of one sensor, i.e., $N=1$. Suppose $L=2,~K=3,~\rho = 2$, and two sets of parameters $\{c_1,c_2\}$ in \eqref{alpha} are $c_1^{(a)}=0.4,~c_2^{(a)}=1$ and $c_1^{(b)}=0.8,~c_2^{(b)}=1$. Since $K=3$, the battery state process is a 4-state Markov chain. Fig. (\ref{fig_battery}) is the schematic representation of this 4-state Markov chain. We consider two sets of parameters $\{\mu_{1,1}, \theta_1\}$, $\mu_{1,1}^{(a)}=1.5, \theta_1^{(a)} = 3$ and $\mu_{1,1}^{(b)} = 1, \theta_1^{(b)}=1.5$. The corresponding $4 \times 4$ transition matrices, denoted as $\Psi_1^{(a)}$ and $\Psi_1^{(b)}$, are
\vspace{-2mm}
\begin{align}\label{psi1}
\Psi_1^{(a)} = 
\begin{pmatrix}
0.0183 & 0.0733 & 0.1465 & 0.7619\\    
0.0001  & 0.0185 & 0.0735  & 0.9080\\           
0.0001  & 0.0002 & 0.0187 & 0.9810\\          
0.0001 & 0.0002 & 0.0057 & 0.9940\\            
\end{pmatrix},\nonumber
\\\Psi_1^{(b)}=
\begin{pmatrix}
0.0276 & 0.0743 & 0.1463 & 0.7518\\        
0.0011 & 0.0216 & 0.0776 &  0.8997\\         
0.0011 & 0.0114 &0.0470 & 0.9406\\        
0.0011 & 0.0114 & 0.0367 & 0.9508\\          
\end{pmatrix}.
\end{align}
 Fig. (\ref{alpha_map}) illustrates the corresponding transmission symbol $\alpha_{1,t}$, i.e., the map in \eqref{alpha} shows how many energy units the sensor should spend for its data transmission, given which of the two quantization intervals $g_{1,t}$ belongs to and its battery state. For instance, for the parameters in Fig. (\ref{powermap1}), when $g_{1,t} \in \mathcal{I}_{1,1}$ and $B_{1,t}=3$, then $\alpha_{1,t}=1$, whereas for the parameters in Fig. (\ref{powermap2}), when $g_{1,t} \in \mathcal{I}_{1,1}$ and $B_{1,t}=3$, then $\alpha_{1,t}=2$.
\begin{figure}[!t]
\begin{subfigure}[t]{0.5\textwidth}
  \centering
  \includegraphics[scale=.47]{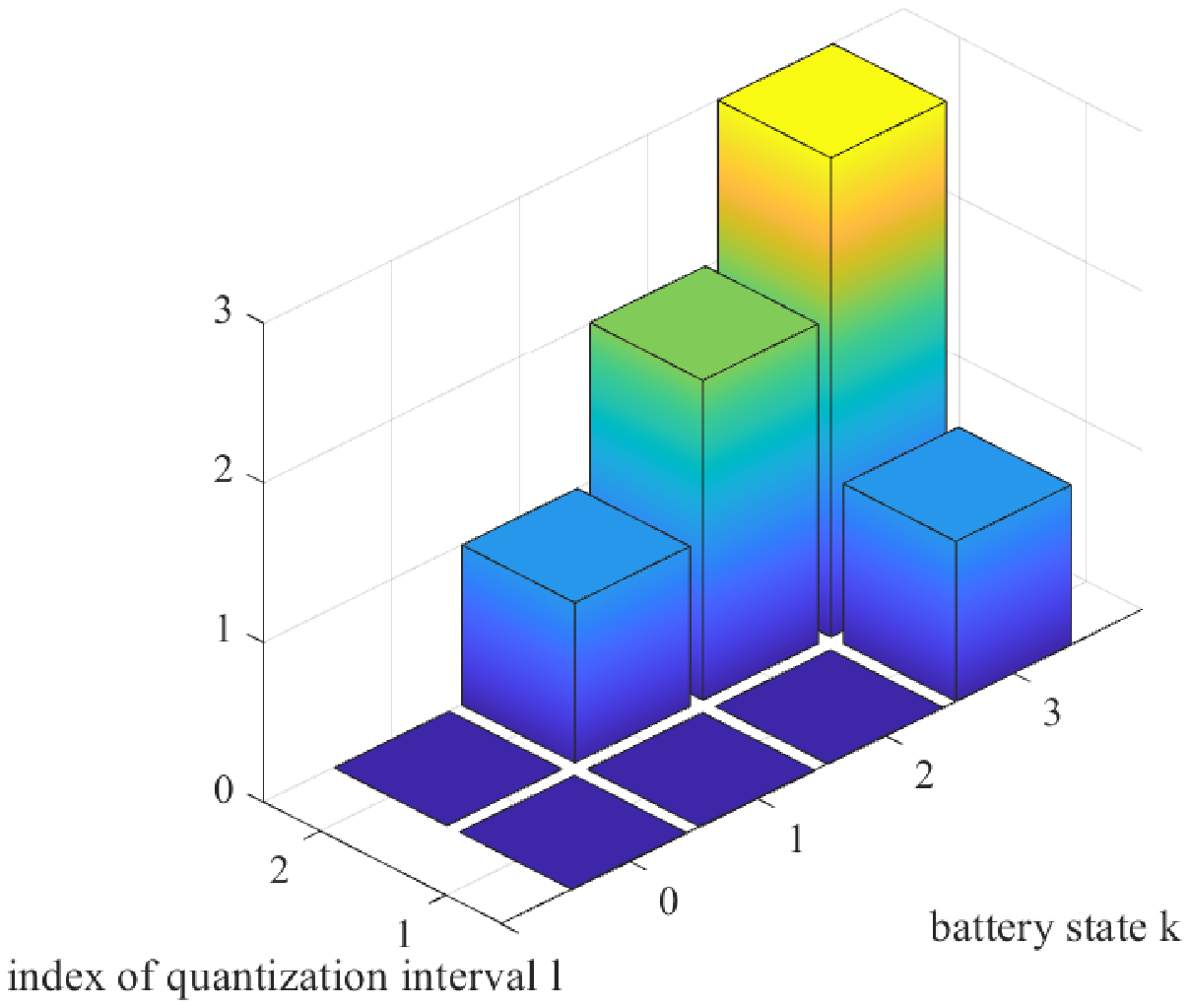}  
  \caption{{$\mu_{1,1}^{(a)}=1.5, \theta_1^{(a)} = 3,$ $c_1^{(a)}=0.4, c_2^{(a)}=1$}}
 \label{powermap1}
\end{subfigure}
\begin{subfigure}[t]{0.5\textwidth}
  \centering
  \includegraphics[scale=.47]{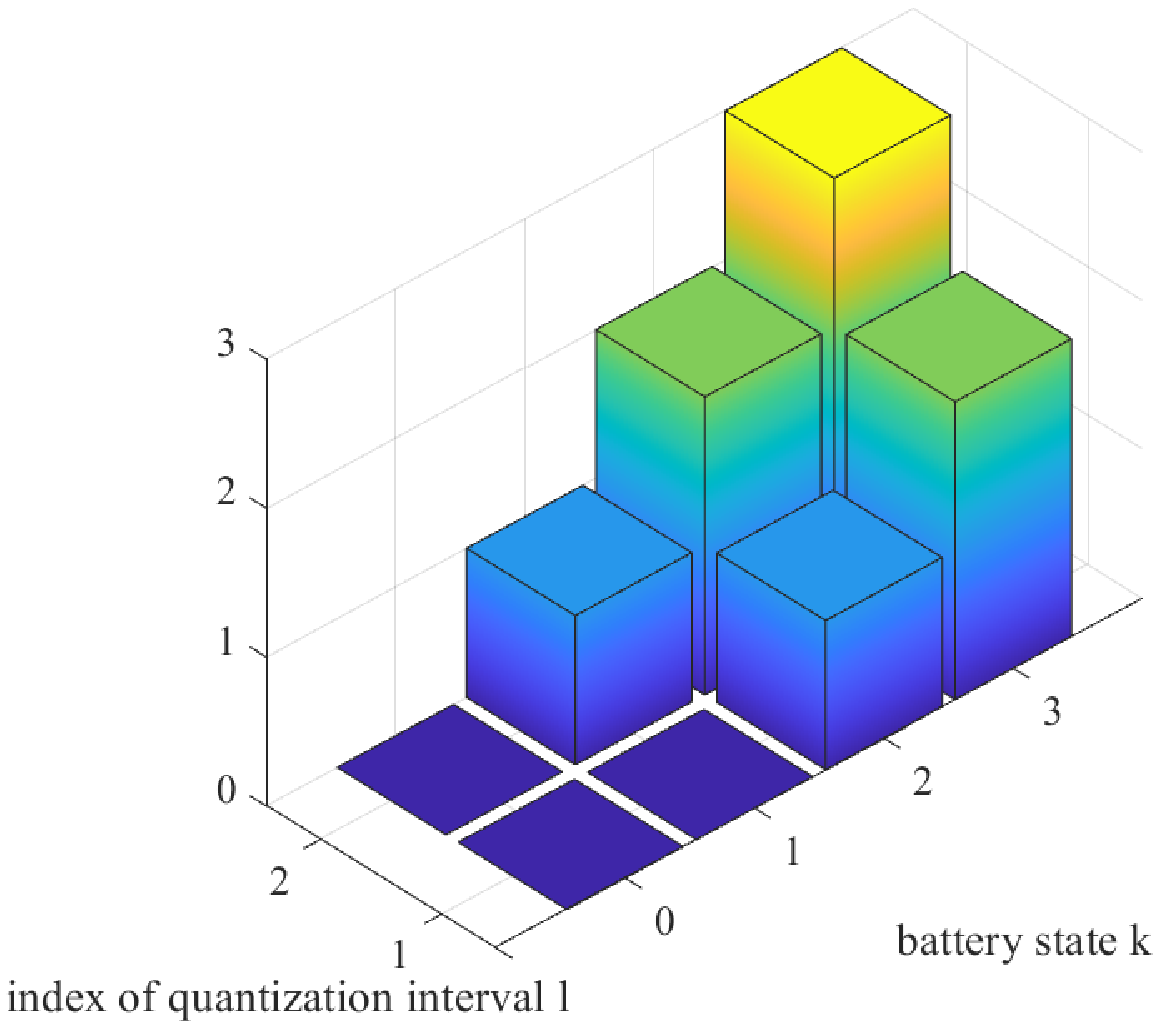}  
  \caption{$\mu_{1,1}^{(b)} = 1, \theta_1^{(b)}=1.5,$ $c_1^{(b)}=0.8, c_2^{(b)}=1$}
  \label{powermap2}
\end{subfigure}
\caption{This numerical example shows how many energy units $\alpha_{1,t}$ the single sensor  should spend for its data transmission, given the knowledge of its channel gain and battery state.}
\label{alpha_map}
\end{figure}
 \subsection{Received Signals at FC and Optimal Bayesian Fusion Rule}
In each time slot sensors send their data symbols to the FC over orthogonal fading channels.   
 The received signal at the FC from sensor $n$ corresponding to time slot $t$ is
\begin{equation}\label{y_n,t}
   y_{n,t} = g_{n,t}\, \alpha_{n,t} + w_{n,t}, ~~~~\text{for}~n=1,\dots,N
\vspace{-1mm}
\end{equation}
where $w_{n,t} \sim {\cal N} (0,\sigma_{w_n}^2)$ is the additive Gaussian noise. We assume $w_{n,t}$'s are i.i.d. over time slots and independent across sensors. 
 Let $\boldsymbol{y}_t=[y_{1,t},y_{2,t}, \ldots ,y_{N,t}]$ denote the vector that includes the received signals at the FC from all sensors in time slot $t$. The FC applies the optimal Bayesian fusion rule $\Gamma_0(.)$ to the received vector $\boldsymbol{y}_t$ and obtains a global decision $u_{0,t}=\Gamma_0(\boldsymbol{y}_t)$, where $u_{0,t} \in \{0,1\}$ \cite{ahmadi}. In particular, we have
 \begin{equation}\label{u0_lrt}
 u_{0,t}= \Gamma_0(\boldsymbol{y}_t)= \\
 \begin{cases}
 1,~~~~~~\Delta_t > \tau, \\
 0,~~~~~~ \Delta_t < \tau,
 \end{cases}
\end{equation}
 where the decision threshold $\tau =\log( \frac{\Pi_0}{\Pi_1})$ and 
\begin{equation}\label{lrt1}
\Delta_t=\log\left (\frac{f(\boldsymbol{y}_t|h_t=1)}{f(\boldsymbol{y}_t|h_t=0)}\right),
\end{equation}
and $f(\boldsymbol{y}_t|h_t)$ is the conditional probability density function (pdf) of the received vector $\boldsymbol{y}_t$ at the FC. 
 \subsection{Our Proposed Constrained Optimization Problems}
From Bayesian perspective, the natural choice to measure the detection performance corresponding to the global decision $u_{0,t}$ at the FC is the error probability, defined as 
\begin{align}\label{pe1}
 P_e &=\Pi_0 \Pr (u_{0,t}=1|h_t=0)+\Pi_1 \Pr(u_{0,t}=0|h_t=1)\nonumber\\
&=\Pi_0 \Pr(\Delta_t > \tau |  h_t=0) + \Pi_1 \Pr(\Delta_t < \tau |  h_t=1).   
\end{align}
 However, finding a closed form expression for $P_e$ is often mathematically intractable. 
Instead, we choose the $J$-divergence between the distributions of the detection statistics at the FC under different hypotheses, as the detection performance metric. This choice allows us to provide a more tractable analysis. We consider two constrained optimization problems. In the first problem $(\mathcal{P}_1)$, we maximize the average total $J$-divergence, subject to an average transmit power per sensor constraint. In the second problem $(\mathcal{P}_2)$, we minimize the average total transmit power, subject to an average $J$-divergence per sensor constraint. In other words, we are interested in solving the following constrained optimization problems
\begin{align}\label{Eg1}
(\mathcal{P}_1):& \max_{\{\theta_n,\mu_{n,l}\},{\forall n,l}} ~\text{average total \textit{J}-divergence}\nonumber\\
   &{\hbox{s.t.}}~\text{average transmit power per sensor}\leq \alpha_{0},~ \forall n
\end{align}
and
\begin{align}\label{Eg2}
(\mathcal{P}_2):& \min_{\{\theta_n,\mu_{n,l}\},{\forall n,l}}\text{average total transmit power}~~~~\nonumber\\
   & {\hbox{s.t.}}~\text{average \textit{J}-divergence per sensor}\geq J_{0}, ~\forall n
\end{align}
where the optimization variables are the set of local thresholds $\{\theta_n\}_{n=1}^N$ and the set of the quantization thresholds $\{\mu_{n,l}\}_{l=1,n=1}^{L,N}$.
\ Note that, from this point forward, we assume that the battery operates at its steady state and we drop the superscript $t$.
\section{Characterization of \textit{J}-divergence and Error Probability}\label{J_error}
In this section, first we define the total $J$-divergence and then we derive a closed-form expression for it in Section III.A, using Gaussian approximation. Next, we attempt to approximate $P_e$ in \eqref{pe1} and provide two closed-form approximate expressions for it in Section III.B. 
\subsection{\textit{J}-Divergence Derivation}\label{j_drive}
We start with the definition of $J$-divergence. Consider two pdfs of a continuous random variable $x$, denoted as $\eta_1(x)$ and $\eta_2(x)$. By definition \cite{vin}, \cite{zahra}, the $J$-divergence between $\eta_1(x)$ and $\eta_0(x)$, denoted as $J(\eta_1,\eta_0)$, is
\begin{equation}\label{first_j}
    J(\eta_1,\eta_0) = D(\eta_1||\eta_0)+D(\eta_0||\eta_1),
\end{equation}
where $D(\eta_i||\eta_j)$ is the non-symmetric Kullback-Leibler (KL) distance between $\eta_i(x)$ and $\eta_j(x)$. The KL distance $D(\eta_i||\eta_j)$ is defined as
\begin{equation}\label{kl}
    D(\eta_i||\eta_j) = \int_{-\infty}^{\infty} \log \left(\frac{\eta_i(x)}{\eta_j(x)}\right)\eta_i(x) dx.
\end{equation}
Substituting \eqref{kl} into \eqref{first_j} we obtain
\begin{equation}\label{j23}
J(\eta_1,\eta_0)=\int_{-\infty}^{\infty} \left[\eta_1 (x)-\eta_0 (x)\right] \log \left(\frac{\eta_1(x)}{\eta_0(x)}\right) dx. 
\end{equation}
In our problem setup, the two conditional pdfs $f(\boldsymbol{y} |h=1)$ and $f(\boldsymbol{y} |h=0)$ play the role of $\eta_1(x)$ and $\eta_0(x)$, respectively. Let $J_{tot}$ denote the $J$-divergence between $f(\boldsymbol{y}| h=1)$ and  $f(\boldsymbol{y}|h=0)$. The pdf of vector $\boldsymbol{y}$ given $h$ is
\begin{align}\label{y/H}
f(\boldsymbol{y}|h)&\overset{(a)}=\prod_{n=1}^N f(y_n|h)   \nonumber\\
& \overset{(b)}= \prod_{n=1}^N f(y_{n}|\alpha_{n}, h)\Pr(\alpha_{n}|h)
\nonumber\\&\overset{(c)}=\prod_{n=1}^N \underbrace{f(y_{n}|\alpha_{n})\Pr(\alpha_{n}|h)}_{=f(y_{n}|h)}, ~~~~~~\text{for}~h=0,1.
\end{align}
Equality ($a$) in \eqref{y/H} holds since the received signals from sensors at the FC, given $h$, are conditionally independent, equality ($b$) in \eqref{y/H} is obtained from Bayes' rule, and equality ($c$) in \eqref{y/H} is found noting that $H$, $\alpha_n$, $y_n$ satisfy the Markov property, i.e.,  $H \rightarrow \alpha_n \rightarrow y_n$ \cite{vin}, \cite{zahra} and hence $y_n$ and $H$, given $\alpha_n$, are conditionally independent. 
Let $J_n$ represent the $J$-divergence between the two conditional pdfs $f({y}_n|h=1)$ and $f({y}_n|h=0)$. Using \eqref{j23} we can express $J_n$ as
\begin{align}
    J_n = &\nonumber\\
   \int_{-\infty}^{\infty}& \!\!\Big [f(y_n| h=1)\!-\! f(y_n| h=0)\Big ] {\rm log}\left({\frac{f(y_n| h=1)} {f(y_n|h=0)}}\right)\;dy_n.
\end{align}
Based on \eqref{y/H} we have $J_{tot}= \sum_{n=1}^N J_n.$
To calculate $J_n$, we need to find the conditional pdf $f(y_n|h)$. Considering \eqref{y_n,t} we realize that $y_n$, given $\alpha_n$, is Gaussian.
In particular, we have
\begin{equation}\label{Gaussian-dis}
f(y_n | \alpha_n=0)= \mathcal{N}(0, \sigma_{w_n}^2), f(y_n |\alpha_n \neq 0)=\mathcal{N}(g_n \alpha_n, \sigma_{w_n}^2)
\end{equation}
Also, considering \eqref{pd_pf1} we find
\begin{align}\label{Gaussian-dis2}
  &\Pr(\alpha_n\neq 0|h=0)= P_{\text{f}_n},~
  \Pr(\alpha_n \neq 0 |h=1)= P_{\text{d}_n},\\
  &\Pr(\alpha_n = 0 |h=0) = 1\!-\!P_{\text{f}_n},~ \Pr(\alpha_n = 0|h=1) = 1\!-\! P_{\text{d}_n}.\nonumber  
\end{align}
Substituting \eqref{Gaussian-dis} and \eqref{Gaussian-dis2} in \eqref{y/H}, the conditional pdfs $f(y_n|h=0)$ and $f(y_n|h=1)$ become
\begin{align}
\label{xyz}
&f(y_n |h=0)=f(y_n |\alpha_n \neq 0) P_{\text{f}_n} + f(y_n |\alpha_n = 0) (1-P_{\text{f}_n}),\nonumber\\
&f(y_n |h=1)=f(y_n |\alpha_n \neq 0) P_{\text{d}_n} + f(y_n |\alpha_n = 0) (1-P_{\text{d}_n}).
\end{align}
Although $f(y_n |\alpha_n=0)$ and $f(y_n | \alpha_n \neq 0)$ in \eqref{xyz} are Gaussian, $f(y_n|h=0)$ and $f(y_n|h=1)$ are Gaussian mixtures, due to $P_{\text{d}_n}$ and $P_{\text{f}_n}$. 
Unfortunately, the $J$-divergence between two Gaussian mixture densities does not have a general closed-form expression. Similar to \cite{vin}, \cite{zahra} we approximate the $J$-divergence between two Gaussian mixture densities by the $J$-divergence between two Gaussian densities $f^G(y_n|h) \sim {\cal N}(m_{n,h},\Upsilon_{n,h}^2)$, where the mean $m_{n,h}$ and the variance $\Upsilon_{n,h}^2$ of the approximate distributions are obtained from matching the first and second order moments of the actual and the approximate distributions. For our problem setup, one can verify that the parameters $m_{n,h}$ and $\Upsilon_{n,{h}}^2$ become
\begin{align}\label{G_par}
    &m_{n,0}={{g_n}\alpha_n}P_{\text{f}_n}\nonumber,~~~ \Upsilon_{n,0}^2\!=\!g_n^2\alpha_n^2 P_{\text{f}_n}(1\!-\!P_{\text{f}_n})\!+\!\sigma_{w_n}^2,\\ 
&m_{n,1}\!=\!{{g_n}\alpha_n}P_{\text{d}_n},~~~~ \Upsilon_{n,1}^2\!=\!g_n^2\alpha_n^2 P_{\text{d}_n}(1\!-\!P_{\text{d}_n})\!+\!\sigma_{w_n}^2.
\end{align}
The $J$-divergence between two Gaussian densities, represented as $J\big(f^G(y_n|h=1),f^G(y_n|h=0)\big)$, in terms of their means and variances is \cite{vin}
\begin{align}\label{g_J}
    & J\big(f^G(y_n|h=1),f^G(y_n| h=0)\big)= \nonumber\\
     &\frac{\Upsilon_{n,1}^2\!+\!( m_{n,1}\!-\! m_{n,0})^2}{\Upsilon_{n,0}^2}
     +\frac{\Upsilon_{n,0}^2\!+\!( m_{n,0}\!-\! m_{n,1})^2}{\Upsilon_{n,1}^2}. 
\end{align}
Substituting $m_{n,h}$ and $\Upsilon_{n,h}^2$ into $J_n$ in \eqref{g_J} we approximate $J_n$ as the following\vspace{5mm}
\begin{align}\label{sim_j}
    J_n=\frac{\sigma_{w_n}^2 + A_n g_n^2\alpha_n^2}{\sigma_{w_n}^2 + B_n g_n^2\alpha_n^2}&+\frac{\sigma_{w_n}^2  + C_n g_n^2\alpha_n^2}{\sigma_{w_n}^2  +  D_n g_n^2\alpha_n^2},
\end{align}
 where
\begin{align*} 
A_n =&~P_{\text{f}_n}(1\!-\!P_{\text{d}_n}) + P_{\text{d}_n}(P_{\text{d}_n}\!-\!P_{\text{f}_n}),\nonumber \\
 C_n=&~P_{\text{d}_n}(1-P_{\text{f}_n}) - P_{\text{f}_n}(P_{\text{d}_n}-P_{\text{f}_n}),\nonumber \\
 B_n =& ~P_{\text{d}_n}(1-P_{\text{d}_n}), ~~D_n = P_{\text{f}_n}(1-P_{\text{f}_n}).
 \end{align*}
\subsection{ Two Error Probability Approximations for the Optimal Bayesian Fusion Rule}\label{approxx}
In this section, we provide two closed-form approximate expressions for $P_e$ in \eqref{pe1}. To obtain the first approximate expression for $P_e$, we expand $\Delta$ in \eqref{lrt1} and approximate $\Delta$ with a Gaussian random variable. This approximation is valid when $\sigma_{w_n}^2 \rightarrow \infty$ and hence, we refer to this as ``low SNR approximation". To find the second approximate expression for $P_e$, we approximate $\Delta$ in \eqref{lrt1} using a similar Gaussian distribution approximation as we conducted in Section \ref{j_drive} and therefore, we refer to this as  ``Gaussian distribution approximation".
\subsubsection{Low SNR Approximation}
We can expand $\Delta$ in
\eqref{lrt1} as the following
\begin{align}
\label{LRT}
\Delta\overset{(d)}=&\log \left(\prod_{n=1}^N \frac{f(y_{n}|h=1)}{f(y_{n}|h=0)}\right)\\
\overset{(e)}= &\sum_{n=1}^N \! \log\!  \left(\frac{P_{\text{d}_n}f(y_{n}|\alpha_{n}\! \neq 0)+(1\! -\! P_{\text{d}_n})f(y_{n}|\alpha_{n}\! =\!  0)}{P_{\text{f}_n}f(y_{n}|\alpha_{n}\! \neq 0)+(1\! -\! P_{\text{f}_n})f(y_{n}|\alpha_{n}\! =\! 0)}\! \right) \nonumber\\ 
\overset{(f)}= &\sum_{n=1}^{N}\!\log\!\left(\frac{P_{\text{d}_n}\text{exp}\big({\frac{-(y_{n}\!-{g_n}\alpha_n)^{2}}{2\sigma^{2}_{w_n}}}\big)\!+\!(1\!-\!P_{\text{d}_n})\text{exp}\big({\frac{-y_{n}^{2}}{2\sigma^{2}_{w_n}}}\big)}{P_{\text{f}_n}\text{exp}\big ({\frac{-(y_{n}\!-{ g_n}\alpha_n)^{2}}{2\sigma^{2}_{w_n}}}\big )\!+\!(1\!-\!P_{\text{f}_n})\text{exp}\big ({\frac{-y_{n}^{2}}{2\sigma^{2}_{w_n}}}\big )}\right)\!\nonumber\\
\overset{(g)}=& \sum_{n=1}^{N}\!\log\!\left(\frac{P_{\text{d}_n}+\!(1\!-\!P_{\text{d}_n})\text{exp}\big(\frac{-(2 y_n g_n\alpha_n-g_n^2\alpha_n^2)}{2\sigma^{2}_{w_n}}\big)}{P_{\text{f}_n}+\!(1\!-\!P_{\text{f}_n})\text{exp}\big(\frac{-(2 y_n g_n\alpha_n-g_n^2\alpha_n^2)}{2\sigma^{2}_{w_n}}\big)}\right)\nonumber .
\end{align}
Equality ($d$) in \eqref{LRT} is inferred from equality ($a$)  in \eqref{y/H}, equality ($e$) in \eqref{LRT} is obtained from substituting $f(y_n|h)$ with \eqref{xyz}, and equality ($f$) in \eqref{LRT} is found via replacing $f(y_n |\alpha_n =0)$ and $f(y_n |\alpha_n \neq 0)$ with \eqref{Gaussian-dis}, and (g) in \eqref{LRT} is obtained after factoring out and canceling the common term $\text{exp}\big({\frac{-(y_{n}\!-{g_n}\alpha_n)^{2}}{2\sigma^{2}_{w_n}}}\big)$ from the numerator and the denominator inside the $\log(.)$. 
In low SNR regime as $\sigma^{2}_{w_n}\!\rightarrow \!\infty$ we can use two approximations $e^{-x} \approx 1-x$ and $\log(1+x) \approx x$ for small $x$ to approximate $\Delta$. In particular, applying the first approximation, we can approximate $\Delta$ in \eqref{LRT} as the following
\begin{align}
\label{dell}
    \Delta &\approx\sum_{n=1}^{N}\!\log\!\left(\frac{P_{\text{d}_n}+(1\!-\!P_{\text{d}_n})(1-\frac{2 y_n g_n\alpha_n-g_n^2\alpha_n^2}{2\sigma^{2}_{w_n}})}{P_{\text{f}_n}+(1\!-\!P_{\text{f}_n})(1-\frac{2 y_n g_n\alpha_n-g_n^2\alpha_n^2}{2\sigma^{2}_{w_n}})}\right)\\
    &=\sum_{n=1}^{N}\!\log\!\left(\frac{1+(P_{\text{d}_n}\!-\!1)(\frac{2 y_n g_n\alpha_n-g_n^2\alpha_n^2}{2\sigma^{2}_{w_n}})}{1+(\!P_{\text{f}_n}\!-\!1)(\frac{2 y_n g_n\alpha_n-g_n^2\alpha_n^2}{2\sigma^{2}_{w_n}})}\right)\nonumber\\
    &=\sum_{n=1}^{N}\Bigg[\log\left(1+(P_{\text{d}_n}\!-\!1)(\frac{2 y_n g_n\alpha_n-g_n^2\alpha_n^2}{2\sigma^{2}_{w_n}})\right)\nonumber\\
    &~~~~~~~~-\log\!\left(1+(P_{\text{f}_n}\!-\!1)(\frac{2 y_n g_n\alpha_n-g_n^2\alpha_n^2}{2\sigma^{2}_{w_n}})\right)\Bigg].\nonumber
\end{align}
Applying the second approximation, we can simplify further  $\Delta$ in \eqref{dell} as below
\begin{align}\label{new_del}
   \Delta &\approx\sum_{n=1}^{N}\left((P_{\text{d}_n}-P_{\text{f}_n})(\frac{2 y_n g_n\alpha_n-g_n^2\alpha_n^2}{2\sigma^{2}_{w_n}})\right). 
\end{align}
After some straightforward manipulations we can rewrite $\Delta$ in \eqref{new_del} as a linear combination of $y_n$'s
\begin{equation}\label{aprox_delta}
    \Delta = -\xi+\sum^{N}_{n=1}\nu_{n}y_{n},
\end{equation}
where 
\begin{align}
\xi&=\sum_{n=1}^{N}g_n^2\alpha_n^2(P_{\text{d}_n}-P_{\text{f}_n})/2\sigma^{2}_{w_n},\nonumber\\
\nu_{n}&={g_n\alpha_n}(P_{\text{d}_n}-P_{\text{f}_n})/\sigma^{2}_{w_n}.
\end{align}
Note that $y_n$ in \eqref{aprox_delta} given $h$ is a Gaussian mixture. Using the Gaussian distribution approximation in Section III.A, we approximate $y_n$ given $h$ with a Gaussian distribution, whose mean $m_{n,h}$ and variance $\Gamma^2_{n,h}$ are given in \eqref{G_par}. Hence, $\Delta$ in \eqref{aprox_delta} becomes a linear combination of Gaussians and thus it is a Gaussian with the mean and the variance given below
\begin{equation}
 \mu_{\Delta|h}=-\xi+\sum_{n=1}^{N}\nu_{n}m_{n,h},~\sigma^{2}_{\Delta|h}=\sum_{n=1}^{N}\nu_{n}^{2}\Upsilon_{n,h}^{2}.
\end{equation}
With these approximations the optimal fusion rule in \eqref{u0_lrt} becomes
\begin{equation}\label{u_del}
u_0= \begin{cases}
1, ~~\Delta> \tau,\\
0, ~~\Delta < \tau,
\end{cases}
\end{equation}
where $\Delta$ is the approximation in \eqref{aprox_delta}.
Therefore, the error probability corresponding to the fusion rule in  \eqref{u_del} can be expressed in terms of $Q$-function
\begin{align}
 P_e \label{snr_pe}
   =\Pi_0Q\left(\frac{\tau-\mu_{\Delta|0}}{\sigma_{\Delta|0}}\right)+\Pi_1\left[1-Q\left(\frac{\tau-\mu_{\Delta|1}}{\sigma_{\Delta|1}}\right)\right].
\end{align}
\subsubsection{Gaussian Distribution Approximation}
In the previous section, we approximated $f(y_n|h)$ with $f^G(y_n|h)={\cal N}(m_{n,h},\Upsilon_{n,h}^2)$, where the mean $m_{n,h}$ and the variance $\Upsilon^2_{n,h}$ of the approximate distribution are provided in \eqref{G_par}. Relying on this Gaussian distribution approximation, we can also approximate the conditional pdf $f(\boldsymbol{y}|h)$. In particular, since the received signals at the FC, conditioned on $h$, are independent across sensors (see (\ref{y/H}-a)), we can approximate $f(\boldsymbol{y}|h)$ with  $f^G(\boldsymbol{y}|h)={\cal N}(\varphi_h,\Lambda_h)$, where $\varphi_h$ and $\Lambda_h$ are the mean vector and the diagonal covariance matrix with elements $m_{n,h}$ and $\Upsilon_{n,h}^2$, respectively.
Using this Gaussian distribution approximation, we can approximate $\Delta$ in \eqref{LRT} as
\begin{align}\label{clt_lrt}
 \Delta\approx&\log \left( \frac{f^G(\boldsymbol{y}|h=1)}{f^G(\boldsymbol{y}|h=0)}\right)\\=&\log  \left(\frac{\sqrt{\det\Lambda_0} \text{exp}\left(-\frac{1}{2}(\boldsymbol{y}-\varphi_1)^T\Lambda_1^{-1}(\boldsymbol{y}-\varphi_1)\right)}{\sqrt{\det\Lambda_1}\text{exp}\left(-\frac{1}{2}(\boldsymbol{y}-\varphi_0)^T\Lambda_0^{-1}(\boldsymbol{y}-\varphi_0)\right)}\right)\nonumber\\=& R \!-\!\frac{1}{2}(\boldsymbol{y}\!-\!\varphi_1)^T\Lambda_1^{-1}(\boldsymbol{y}\!-\!\varphi_1) + \frac{1}{2}(\boldsymbol{y}\!-\!\varphi_0)^T\Lambda_0^{-1}(\boldsymbol{y}\!-\!\varphi_0),\nonumber
\end{align}
where $R=\log \left(\frac{\sqrt{\det\Lambda_0}}{\sqrt{\det\Lambda_1}}\right)$. Since the covariance matrices $\Lambda_0$ and $\Lambda_1$ are diagonal, the approximate expression for $\Delta$ in \eqref{clt_lrt} can be rewritten as
\begin{align}
\label{delta1}
   & \Delta\approx R+ \Delta'_N, ~\Delta'_N = \sum_{n=1}^N z_n, \nonumber\\
   &z_n = \frac{(y_n-m_{n,0})^2}{\Upsilon_{n,0}^2} - \frac{(y_n-m_{n,1})^2}{\Upsilon_{n,1}^2},
\end{align}
With the Gaussian distribution approximation, the optimal fusion rule in \eqref{u0_lrt} can be approximated with
\begin{equation}\label{del_N}
u_0= \begin{cases}
1, ~~\Delta'_N > \tau',\\
0, ~~\Delta'_N < \tau', 
\end{cases}
\end{equation}
where $\Delta'_N$ is given in \eqref{delta1} and $\tau'= 2(\tau-R)$. The error probability corresponding to the fusion rule in \eqref{del_N} is
\begin{equation}\label{Pe-Delta'}
P_e=\Pi_0 \Pr(\Delta'_N > \tau' |h=0)+\Pi_1 \Pr(\Delta'_N < \tau' |h=1).
\end{equation}
To find $P_e$ in \eqref{Pe-Delta'} we need the pdf of $\Delta'_N$ given $h$. We note that $z_n$ in \eqref{delta1} can be rewritten as a quadratic function of $y_n$
\begin{align}
\label{z_n2}
    z_n &=  a y_n^2 + b y_n +c, ~~\mbox{where}\nonumber\\
    a &= \frac{1}{\Upsilon^2_{n,0}}\!\!-\!\frac{1}{\Upsilon^2_{n,1}},~b= \frac{2m_{n,1}}{\Upsilon^2_{n,1}}\!-\!\frac{2m_{n,0}}{\Upsilon^2_{n,0}},~ c=\frac{m_{n,0}^2}{\Upsilon^2_{n,0}}\!-\!\frac{m_{n,1}^2}{\Upsilon^2_{n,1}}.
\end{align}
\begin{figure*}
 \begin{align}
\label{pdf_z}
    f(z_n|h) &= \frac{1}{g(z_n)}\left\{f^G_{y_n|h}\left(\frac{\Upsilon^2_{n,0}\Upsilon^2_{n,1}}{2}g(z_n)\!+\!m_{n,0}\Upsilon^2_{n,1}\!-\!m_{n,1}\Upsilon^2_{n,0}\right)\!+\!f^G_{y_n|h}\left(\frac{\!-\!\Upsilon^2_{n,0}\Upsilon^2_{n,1}}{2}g(z_n)\!+\!m_{n,0}\Upsilon^2_{n,1}\!-\!m_{n,1}\Upsilon^2_{n,0}\right)\right\},\nonumber\\
    g(z_n) &= \frac{2}{\Upsilon_{n,1}\Upsilon_{n,1}}\sqrt{(m_{n,0}-m_{n,1})^2+z_n(\Upsilon^2_{n,1}-\Upsilon^2_{n,0})}.
 \end{align}
 \hrulefill
\end{figure*}
Let $\mu_{z_n|h}$ and $\sigma^2_{z_n|h}$, denote the mean and variance of $z_n$ in \eqref{z_n2} given $h$, respectively. 
To find $\mu_{z_n|h}, \sigma^2_{z_n,h}$ we recall the following fact.\par \textbf{Fact}: Let $x \sim N (\mu, \sigma^2)$ be a Gaussian random variable with the mean $\mathbb{E}\{x\}=\mu$ and the variance $\sigma^2=\mathbb{E}\{x^2\}-\mu^2$. Then we have \cite{xx}:
\begin{align}
    \mathbb{E}\{x^2\} &= \mu^2+\sigma^2,\\
    \mathbb{E}\{x^3\} &=\mu(\mu^2+3\sigma^2),\nonumber\\
    \mathbb{E}\{x^4\}&=\mu^4+6\mu^2\sigma^2+3\sigma^4.\nonumber
\end{align}
Using this fact, we find
\begin{align}
&\mu_{z_n|h}=a(m_{n,h}^2+\Upsilon^2_{n,h})+b\,m_{n,h}+c, \\
&\sigma^2_{z_n|h}= 2 a^2( 2 m_{n,h}^2+\Upsilon^4_{n,h})+b \Upsilon^2_{n,h}(b+4\,a\,m_{n,h}),\nonumber
\end{align}
where $a,b,c$ are given in \eqref{z_n2} and $m_{n,h}, \Upsilon^2_{n,h}$ are given in \eqref{G_par}.
Relying on the Gaussian distribution approximation of $y_n$ given $h$, we can derive the pdf of $z_n$ given $h$, where the pdf expression is provided in  \eqref{pdf_z}. Since given $h$, $z_n$'s are independent, the pdf of $\Delta'_N$ given $h$, is convolution of these $N$ individual pdfs, which does not have a closed-form expression. This indicates that, even with the Gaussian distribution approximation, finding a closed-form expression of $P_e$ in \eqref{Pe-Delta'} for finite $N$ remains elusive. Hence, we resort to the asymptotic regime when $N$ grows very large and invoke the central limit theorem (CLT) to approximate $P_e$ in \eqref{Pe-Delta'}.
\par Lindeberg CLT is a variant of CLT, where the random variables are independent, but not necessarily identically distributed \cite{lin_1}. Let $\mu_{\Delta'_N|h}$ and $\sigma^2_{\Delta'_N|h}$ indicate the mean and variance of $\Delta'_N$ in \eqref{delta1} given $h$. We have $\mu_{\Delta'_N|h} = \sum_{n=1}^N  \mu_{z_n|h} $ and $\sigma_{\Delta'_N|h}^2 = \sum_{n=1}^N \sigma^2_{z_n|h}$. Assuming Lindeberg's condition, given below, is satisfied
\begin{equation}
\lim_{N\rightarrow\infty}\frac{1}{\sigma_{\Delta'_N|h}^2}\sum_{n=1}^{N}\mathbb{E}\{(z_n - \mu_{z_n|h})^2\}=0,
\end{equation}
then, as $N$ goes to infinity, the normalized sum $(1/\sigma_{\Delta'_N|h}^2 ) \sum_{n=1}^N(z_n - \mu_{z_n|h})$ converges in distribution toward the standard normal distribution
\begin{equation}\label{lindeberg-CLT}
(1/\sigma_{\Delta'_N|h}^2) \sum_{n=1}^N (z_n - \mu_{z_n|h}) \overset{d}\rightarrow \mathcal{N}(0,1),
\end{equation}
where $\overset{d}\rightarrow$ indicates convergence in distribution. Using \eqref{lindeberg-CLT} we can approximate $P_e$ in \eqref{Pe-Delta'} using $Q$-function
\begin{equation}
\label{clt_pe}
     P_e=\Pi_0Q\left(\frac{\tau'- \mu_{\Delta'_N|0}}{\sigma^2_{\Delta'_N|0}}\right)+\Pi_1\!\left[1\!-\!Q\left(\frac{\tau'-\mu_{\Delta'_N|1}}{\sigma^2_{\Delta'_N|1}}\right)\right].
\end{equation}
\section{Cost Functions in Problems (${\cal P}_1$) and (${\cal P}_2$)}\label{cost_fun}
 In this section, we formulate problems $({\cal P}_1)$ in \eqref{Eg1} and $({\cal P}_2)$ in \eqref{Eg2}, and derive the cost functions and the constraints in terms of the optimization variables.
 \par Recall total $J$-divergence  $J_{tot}=\sum_{n=1}^N J_n$, where $J_n$ in given in \eqref{sim_j}, and transmit power per sensor $\alpha_n^2$ is given in \eqref{alpha}. We note that $J_n$ depends on $g_n$ value, whereas $\alpha_n^2$ depends on the quantization interval to which $g_n$ belongs. The dependency of $J_n$ on $g_n$ stems from the fact that the FC is equipped with a coherent receiver with full knowledge of all channel gains $g_n$'s, and the optimal Bayesian fusion rule utilizes this collective information. Hence, the error probability $P_e$ and its bound $J_{tot}$ depend on this information. On the other hand, sensor $n$ only knows the  quantization interval to which $g_n$ belongs, and adapts its transmit power $\alpha_n^2$ according to this partial channel state information (as well as its battery state). 
\par Furthermore, the cost function and the constraint in both $({\cal P}_1)$ and $({\cal P}_2)$ problems are decoupled across sensors. Hence, both optimization problems can lend themselves to a distributed implementation, where each sensor can optimize locally its optimization variables, given the locally available information. In distributed implementation, sensor $n$ solves for the local threshold $\theta_n$ and the quantization thresholds $\{\mu_{n,l}\}_{l=1}^L$. The FC also solves for the local threshold $\theta_n$ and the quantization thresholds $\{\mu_{n,l}\}_{l=1}^L$ for all sensors. Employing these optimized variables, the FC quantizes $g_n$'s and provides each sensor (via a low-rate feedback channel) with the information on the corresponding index of quantization interval. Given the optimized variables for $\theta_n$ and $\{\mu_{n,l}\}_{l=1}^L$, as well as the received feedback information and its battery state, sensor $n$ chooses its transmission power $\alpha_n^2$ according to \eqref{alpha}. To enable such a distributed implementation, we should take the average of $J_n$ and $\alpha_n^2$ over $g_n$, conditioned that $g_n \in [\mu_{n,l},\mu_{n,l+1}]$. By taking the conditional average over $g_n$, both $({\cal P}_1)$ and $({\cal P}_2)$ problems can be solved offline and the solutions we obtain are valid, as long as the channel gain statistics remain unchanged\footnote{We note that without taking the average over $g_n$ (conditioned that $g_n$ lies in a specified interval),  the solution of the optimization problem cannot have a distributed implementation, since sensor $n$ does not have full knowledge of $g_n$ and only knows the quantization interval to which $g_n$ belongs.}.\par Let $\bar{J}_n^{(i)}=\mathbb{E}\{ J_n | g_n \in [\mu_{n,i}, \mu_{n, i+1} ]\}$ and $ \bar{\alpha}_n^{(i)}=\mathbb{E}\{\alpha_n | g_n \in [\mu_{n,i}, \mu_{n, i+1}] \}$, respectively, denote the expectations of $J_n$ and $\alpha_n$ over $g_n$ and, conditioned that $g_n \in [\mu_{n,i}, \mu_{n,i+1}]$.
In the following, we compute the two conditional expectations $\bar{J}_n^{(i)}$ and $\bar{\alpha}_n^{(i)}$, in terms of the optimization variables $\{\theta_n\}_{n=1}^N$ and $\{\mu_{n,l}\}_{l=1,n=1}^{L,N}$. To compute $\bar{J}_n^{(i)}$ we use the following fact.
\par
{\bf Fact}: Suppose random variable $x$ has an exponential distribution with parameter $\lambda$, i.e., $f(x)=\lambda e^{-\lambda x}$. Consider the function $h(x) = \frac{a + b x}{c + d x}$, with given constants $a$, $b$, $c$ and $d$. Then, the average of $h(x)$, conditioned on $x$ being in the interval $[\mu_i,\mu_{i+1}]$ is
\begin{align}\label{ex1}
  &\mathbb{E}\{h(x) | x \in [\mu_i,\mu_{i+1}]\}=\int_{\mu_i}^{\mu_{i+1}} h(x)f(x)dx\nonumber\\ 
    &= \frac{1}{d}\Big[a\beta(\mu_{i+1})-\frac{bc}{d}\beta(\mu_{i+1})-be^{-\lambda \mu_{i+1}}\nonumber\\ 
    &~~~~~~~~~~~~~~~~-a\beta(\mu_{i})-\frac{bc}{d}\beta(\mu_{i})-be^{-\lambda \mu_{i}}\Big],
\end{align}
where
 \begin{align}\label{ex2}
     \beta(x) &=\lambda \text{exp}{(\frac{c\lambda}{d})}~ \text{Ei} \Big(-\lambda x - \frac{c\lambda}{d}\Big),\\
     \text{Ei}(z)&=\int^{\infty}_{-z} \frac{e^{-t}}{t} dt.\nonumber
 \end{align}
Using this fact and letting $a_1=a_2=c_1=c_2=\sigma^2_{w_n}$, $b_1=A_n \alpha_n^2,~ b_2= C_n \alpha_n^2$, $d_1=B_n \alpha_n^2$ and $d_2=D_n \alpha_n^2$, where $A_n, B_n, C_n, D_n$ are given in \eqref{sim_j},  we reach at
\begin{align}\label{ave_J}
  \bar{J}_n^{(i)} = \sum_{k=0}^K\phi_{n,k}\pi_{n,i}\Big[\Omega(\lfloor c_i k\rfloor,\mu_{n,i+1}^2)-\Omega(\lfloor c_i k\rfloor,\mu_{n,i}^2)\Big],
\end{align}
where the two dimensional function $\Omega(x,y)$ in \eqref{ave_J} is
\begin{align*}
 &\Omega(x,y)\triangleq
  \nonumber\\&\frac{1}{B_n x}\Big[\sigma^2_{w_n}\beta_1(x,y)-\frac{A_n}{B_n}\sigma^2_{w_n}\beta_1(x,y) - A_n x  e^{(-y\gamma_{g_n} )}\Big]+
  \nonumber\\&
  \frac{1}{D_n x}\Big[\sigma^2_{w_n}\beta_2(x,y) - \frac{C_n}{D_n}\sigma^2_{w_n}\beta_2(x,y) - C_n x e^{(-y\gamma_{g_n} )}\Big],\nonumber\\
\end{align*}
and

\begin{align*}
&\beta_1(x,y)\triangleq\gamma_{g_n}\text{exp}\Big(\frac{\sigma^2_{w_n}\gamma_{g_n}}{x B_n} \Big)\text{Ei}\Big(-\gamma_{g_n}y-\frac{\sigma^2_{w_n}\gamma_{g_n}}{x B_n}\Big),\nonumber
\end{align*}
\begin{align*}
&\beta_2(x,y)\triangleq\gamma_{g_n}\text{exp}\Big(\frac{\sigma^2_{w_n}\gamma_{g_n}}{x D_n}\Big)\text{Ei}\Big(-\gamma_{g_n}y-\frac{\sigma^2_{w_n}\gamma_{g_n}}{x D_n}\Big). 
\end{align*}
We can compute $\bar{\alpha}_n^{(i)}$ using \eqref{alpha} as the following
\begin{equation}\label{ave_alph}
    \bar{\alpha}_n^{(i)}=\widehat{\Pi}_{n,1}\sum_{k=0}^{K}\phi_{n,k}\pi_{n,i}\lfloor c_i k\rfloor 
\end{equation}
Then problem $({\cal P}_1)$ in \eqref{Eg1} and problem $({\cal P}_2)$ in \eqref{Eg2} can be formulated as
\begin{align}\label{Eg11}
   ({\cal P}_1)~~&\max_{\{\theta_n,\mu_{n,l}\},{\forall n,l}} \sum_{n=1}^N\sum_{i=0}^L\bar{J}_n^{(i)}\nonumber\\
  &~~~~{\hbox{s.t.}}~\sum_{i=0}^L  \bar{\alpha}_n^{(i)}\leq\alpha_0,~\forall n
\end{align}
and
\begin{align}\label{Eg22}
   {(\cal P}_2)~~ &\min_{\{\theta_n,\mu_{n,l}\},{\forall n,l}}\sum_{n=1}^N\sum_{i=0}^L  \bar{\alpha}_n^{(i)}\nonumber\\
   & {\hbox{s.t.}}~\sum_{i=0}^L\bar{J}_n^{(i)}\geq J_0,~\forall n
\end{align}
\begin{figure}[!t]
\begin{subfigure}[t]{0.5\textwidth}
  \centering
  \includegraphics[scale=.47]{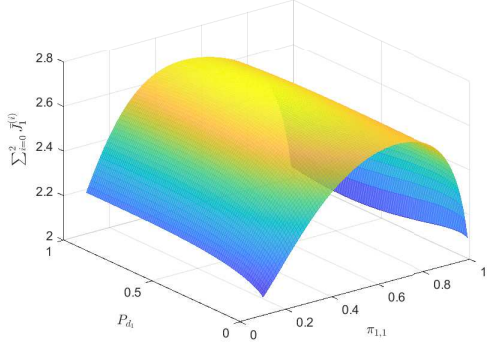}  
  \caption{$\sum_{i=0}^2\bar{J}_1^{(i)}$ vs. $P_{\text{d}_1}$ and $\pi_{1,1}$.}
	\label{ave_j}
\end{subfigure}
\begin{subfigure}[t]{0.5\textwidth}
  \centering
 \includegraphics[scale=.45]{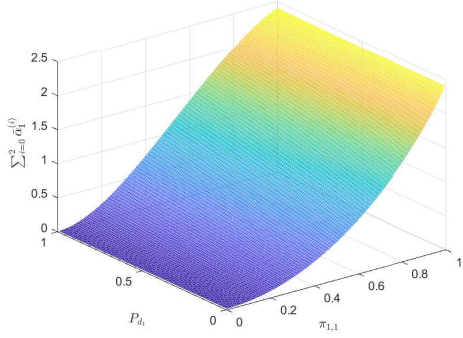}
   \caption{$\sum_{i=0}^2\bar{\alpha}_1^{(i)}$ and vs. $P_{\text{d}_1}$ and $\pi_{1,1}$.}
	\label{ave_alpha}
\end{subfigure}
\caption{$\sum_{i=0}^2\bar{J}_1^{(i)}$ and $\sum_{i=0}^2\bar{\alpha}_1^{(i)}$ vs. $P_{\text{d}_1}$ and $\pi_{1,1}$.}
\end{figure}
\noindent Considering \eqref{alpha}, the dependency of $\bar{\alpha}_n^{(i)}$ on the optimization variables $\theta_n$ and $\{\mu_{n,l}\}_{l=1}^L$ is clear. Examining $\bar{J}_n^{(i)}$, we note that it depends on $\theta_n$, via $P_{\text{d}_n},P_{\text{f}_n}$, and depends on $\{\mu_{n,l}\}_{l=1}^L$ via $\{\pi_{n,l}\}_{l=1}^L$. Problems $(\mathcal{P}_1)$ and $(\mathcal{P}_2)$ do not have closed form analytical solutions and we resort to numerical search methods to solve them. The following observation will help us to significantly reduce the computational complexity of the numerical search. \par Considering \eqref{pd_pf1} we realize that there is a one-to-one correspondence between $\theta_n$ and $P_{\text{d}_n}$. Also, considering \eqref{gaz} we realize that there is a one-to-one correspondence between $\mu_{n,l}$ and $\pi_{n,l}$. Therefore, instead of conducing numerical search to find $\theta_n, \{\mu_{n,l}\}_{l=1}^L$, we can perform numerical search to obtain $P_{\text{d}_n}, \{\pi_{n,l}\}_{l=1}^L$. While the search space for finding  $\theta_n, \{\mu_{n,l}\}_{l=1}^L$ for all sensors is $\mathbb{R}^N \times \mathbb{R}^{N+L}$, the search space for finding $P_{\text{d}_n}, \{\pi_{n,l}\}_{l=1}^L$ for all sensors is $(0,1)^{N+L}$. In Section VI we consider solving problems $(\mathcal{P}_1)$ and $(\mathcal{P}_2)$, where the optimization variables are replaced with $P_{\text{d}_n},  \{\pi_{n,l}\}_{l=1}^L$.
\\A remark on the difference between constrained minimization of the approximate expressions of $P_e$ in \eqref{snr_pe} and \eqref{clt_pe} and problems $(\mathcal{P}_1)$ and $(\mathcal{P}_2)$ follows.
\par {\bf Remark}: Different from $(\mathcal{P}_1)$ and $(\mathcal{P}_2)$, none of the approximate expressions of $P_e$ in \eqref{snr_pe} and \eqref{clt_pe} can be decoupled across sensors. Therefore, constrained minimization of $P_e$ does not render itself to a distributed implementation. \\Mathematically speaking, we are unable to prove that $(\mathcal{P}_1)$ is a concave optimization problem, nor $(\mathcal{P}_2)$ is a convex optimization problem, i.e., the solution to be found by the numerical search method in Section VI is not guaranteed to be the optimal solution (the global maximum in $(\mathcal{P}_1)$ or the global minimum in $(\mathcal{P}_2)$). To gain an insight on the behavior of the cost functions in these problems, we consider the following simple example consisting of one sensor, i.e., $N = 1$. Suppose $L = 2,~K =10,~ \rho = 4$, and the set of parameters $\{c_1,c_2\}$ in \eqref{alpha} is $c_1=0.5$ and $c_2=1$. The original optimization variables $\theta_1, \mu_{1,1}$ are replaced with $P_{\text{d}_1}, \pi_{1,1}$.
Fig.~(\ref{ave_j}) and Fig.~(\ref{ave_alpha}) illustrate $\sum_{i=0}^2\bar{J}_1^{(i)}$ and $\sum_{i=0}^2 \bar{\alpha}_1^{(i)}$ vs. $P_{\text{d}_1}$ and $\pi_{1,1}$. These figure show that $\sum_{i=0}^2\bar{J}_1^{(i)}$ is a jointly convex function of $P_{\text{d}_1}, \pi_{1,1}$ and $\sum_{i=0}^2 \bar{\alpha}_1^{(i)}$ is a jointly concave function of $P_{\text{d}_1}, \pi_{1,1}$. 
\section{Random Deployment of sensors} \label{random_dep}
The signal model in \eqref{xk} is a widely adopted model in the literature of signal (target) detection in a field, when the
signal source is typically modeled as an isotropic radiator (with a general intensity decay model), and the emitted power of the signal source at a reference distance $d_0$ is known \cite{cao}. Suppose $P_0$ is the emitted power of the signal source at  the reference distance $d_0$, and $d_n$ is the Euclidean distance between the source and sensor $n$. For a general intensity decay model, the signal intensity at sensor $n$, denoted as $s_n$,  is \cite{cao}, \cite{niu}
\begin{equation}
    s_n = \frac{P_0}{(d_n/d_0)^\gamma},
\end{equation}
where $\gamma $ is the path-loss exponent, e.g., for free-space wave propagation $\gamma=2$. With this model, the problem of signal detection (when the signal is corrupted by the additive Gaussian noise) is equivalent to the following  binary hypothesis testing problem 
\begin{equation}\label{xs}
H={1}: x_{n} = s_{n}+v_{n}, ~~~~~H={0}: x_{n} = v_{n},
\end{equation}
in which $s_n$ and variance of $v_n$,  denoted as $\sigma_{v_n}^2$,  are assumed to be known \cite{vin}, \cite{maleki1}, \cite{maleki2}. Note that the binary hypothesis testing problem in \eqref{xs} can be recast as the problem in \eqref{xk}, by scaling the sensor observation $x_n$ with $(d_n/d_0)^\gamma$. We note that this signal model applies to any arbitrary, but fixed (given) deployment of sensors in the field. For several WSN applications, the sensors may be deployed randomly in a field. In this case, the sensors' locations are unknown before deployment. This implies that $s_n$ in \eqref{xs} or $\cal A$ in \eqref{xk} are unknown, and consequently, $P_{\text{d}_n}$ and $P_{\text{f}_n}$ in \eqref{pd_pf1} cannot be determined before deployment.
\par To expand our optimization method beyond fixed deployment of sensors, we assume that sensors are randomly deployed in a circle field, the signal source is located at the center of this field, and it is at least $r_0$ meters away from any sensor within the field. Let $r_n$ be the distance of sensor $n$ from the center. We assume $r_n$ is uniformly distributed in the interval $(r_0,r_1)$, i.e.,
\begin{equation}\label{r0}
  f(r_n) =\begin{cases}
  \frac{1}{r_1-r_0},~~r_0<r_n\leq r_1,  \\
  0,~~~~~~~~\text{o.w.}
  \end{cases}
\end{equation}

Suppose the emitted power of the signal source at radius $r_0$ is $P_0$. Then the signal intensity at sensor $n$ is $s_n=\frac{P_0}{(r_n/r_0)^\gamma}$. Given the pdf of $r_n$ in \eqref{r0} and assuming $\gamma=2$, we obtain the pdf of $s_n$ as follows
\begin{equation}\label{omg}
    f(s_n) = 
    \begin{cases}
    \frac{\sqrt{P_0}}{2s_n\sqrt{s_n}(r_1-r_0)},~~~\dfrac{P_0}{r_1^2}<s_n\leq \dfrac{P_0}{r_0^2}, \\
    0,~~~~~~~~~~~~~~~~\text{o.w.}
    \end{cases}
\end{equation}
Based on the pdf of $s_n$ in \eqref{omg} we can recompute $P_{\text{d}_n}$ and $P_{\text{f}_n}$ in \eqref{pd_pf1} as below
\begin{align}\label{pd_pf2}
 &P_{\text{f}_n}=\int_{\frac{P_0}{r_1^2}}^{\frac{P_0}{r_0^2}} \left[Q\Big(\frac{\theta_n+\frac{s_n^2}{2\sigma^2_{v_n}}}{\sqrt{s_n^2/{\sigma^2_{v_n}}}}\Big)\right]f(s_n)d s_n, \nonumber\\
 &P_{\text{d}_n}=\int_{\frac{P_0}{r_1^2}}^{\frac{P_0}{r_0^2}} \left[Q\Big(\frac{\theta_n-\frac{s_n^2}{2\sigma^2_{v_n}}}{\sqrt{s_n^2/{\sigma^2_{v_n}}}}\Big)\right]f(s_n)d s_n.   
\end{align}
We note that, for random deployment of sensors, our proposed constrained optimization problems $(\mathcal{P}_1)$ and $(\mathcal{P}_2)$ are still valid, with the difference that, for $J_n$ in \eqref{sim_j}, $P_{\text{d}_n}$ and $P_{\text{f}_n}$ expressions should be replaced with the ones in \eqref{pd_pf2}.
\section{Simulation results and discussion}\label{simulation}
We corroborate our analysis with Matlab simulations and investigate the following: (i) starting with the steady-state probability vector $\boldsymbol{\Phi}$ in \eqref{inf}, we show how the entries of this probability vector vary, in terms of data transmission and harvesting parameters, (ii) focusing on solving $(\mathcal{P}_1)$, we illustrate the effectiveness of optimizing the parameters $\{\theta_n, \mu_{n,l}\}, \forall n,l$ on enhancing the detection performance of the optimal Bayesian fusion rule at the FC, compared to a system when where these parameters are some arbitrary chosen (not optimized) values, (iii) we examine how the error probability $P_e$ of the optimized system (system with the optimized parameters) changes as different system parameters such as $c_1,~K,~ \rho$ vary, (iv) we examine the accuracy of the two $P_e$ approximations we derived in \eqref{snr_pe} and \eqref{clt_pe}, (v) we examine how the random deployment of sensors affects $P_e$ of the optimized system,
(vi) focusing on solving $(\mathcal{P}_2)$, we show the effectiveness of optimizing the parameters $\{\theta_n, \mu_{n,l}\}, \forall n,l$ in lowering the average total transmit power, compared to a system where these parameters are some arbitrary chosen (not optimized) values, (vii) we examine how the average total transmit power of the optimized system changes as different system parameters change.
\begin{figure}[!t]
\begin{subfigure}[t]{0.24\textwidth}
  \includegraphics[scale=.3]{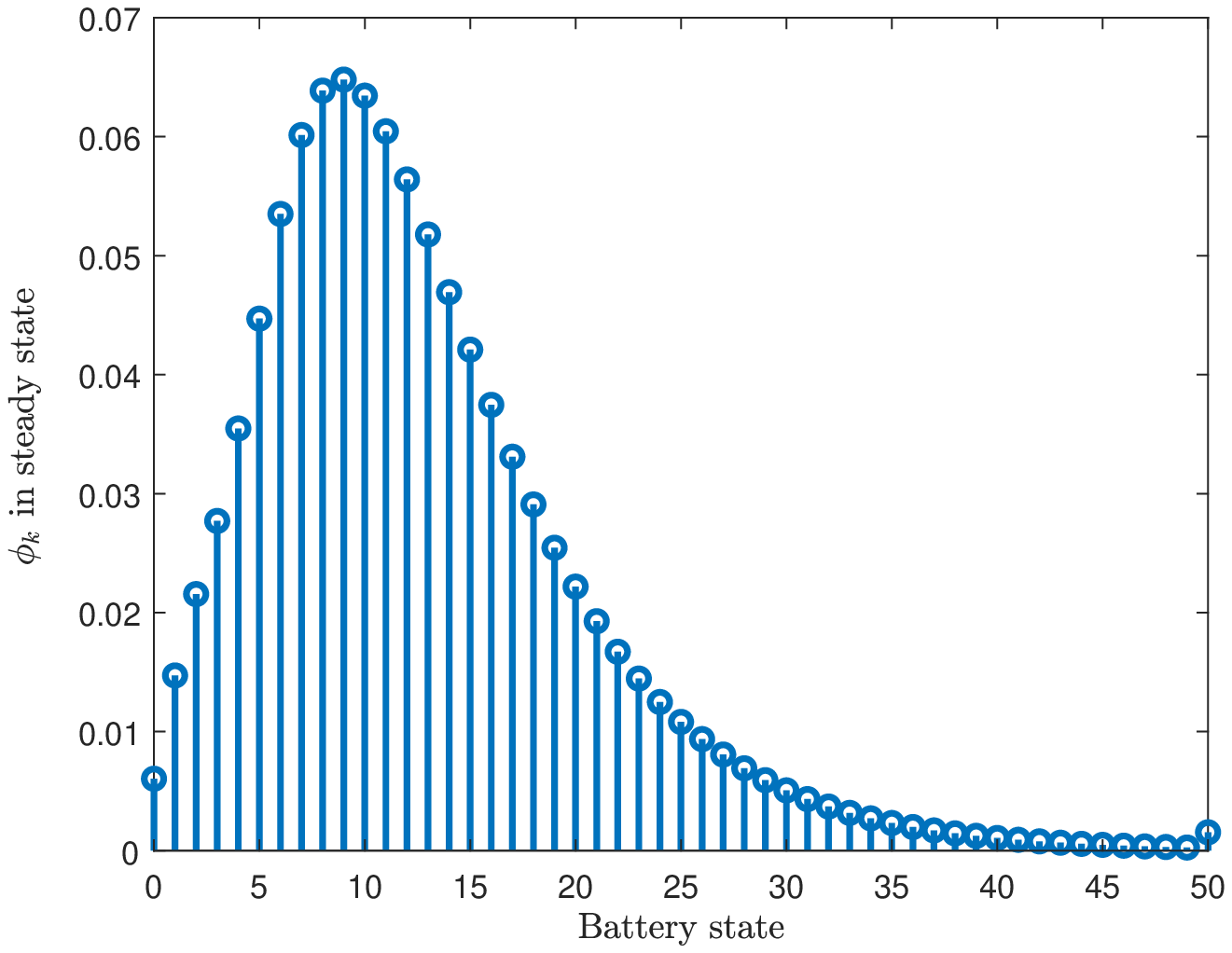}  
  \caption{$\rho=2, c_1 = 0.5$}
 \label{f5a}
\end{subfigure}
\begin{subfigure}[t]{0.24\textwidth}
  \centering
  \includegraphics[scale=.3]{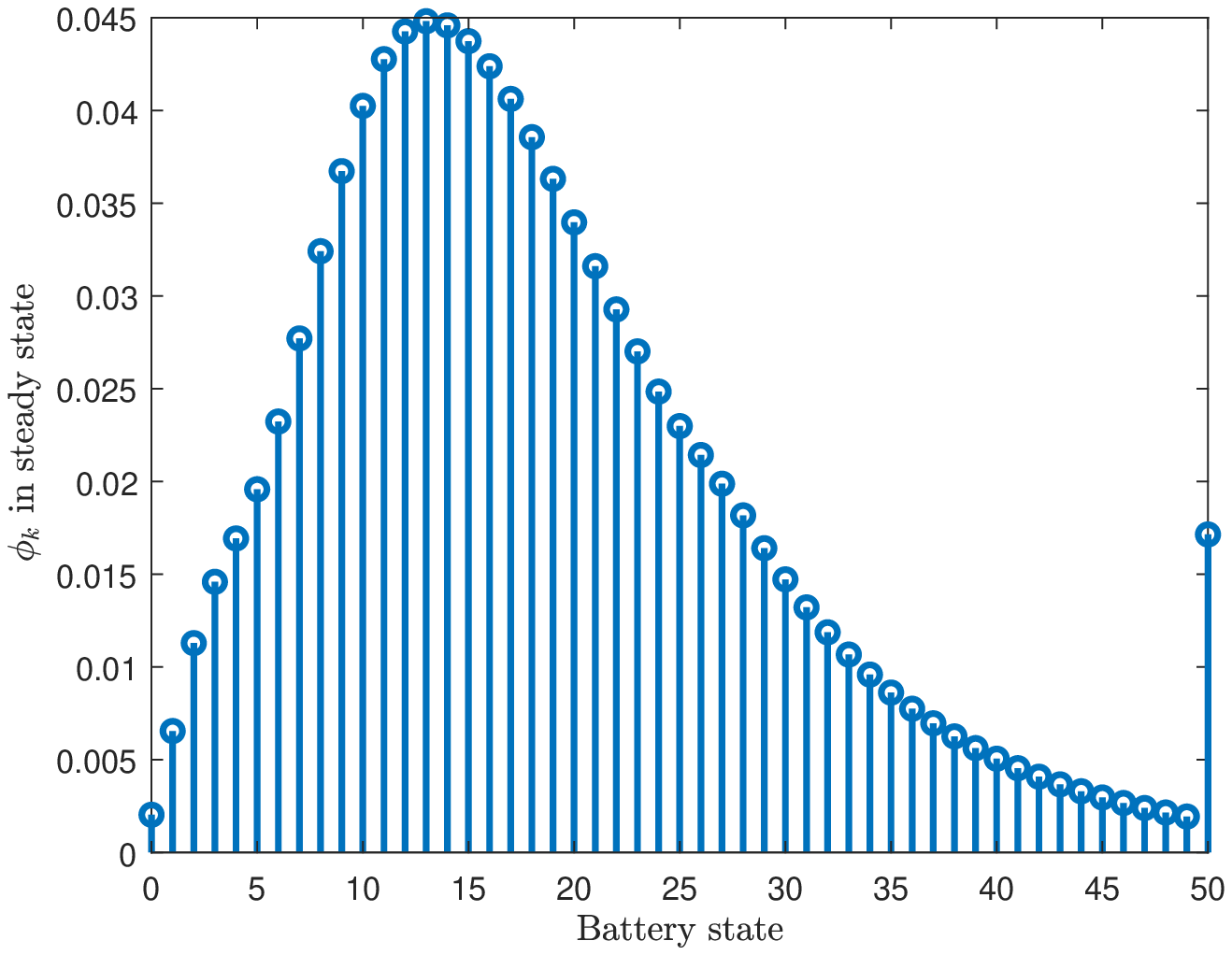}  
  \caption{$\rho=3, c_1 = 0.5$}
  \label{f5b}
\end{subfigure}
\begin{subfigure}[t]{0.24\textwidth}
  \centering
  \includegraphics[scale=.3]{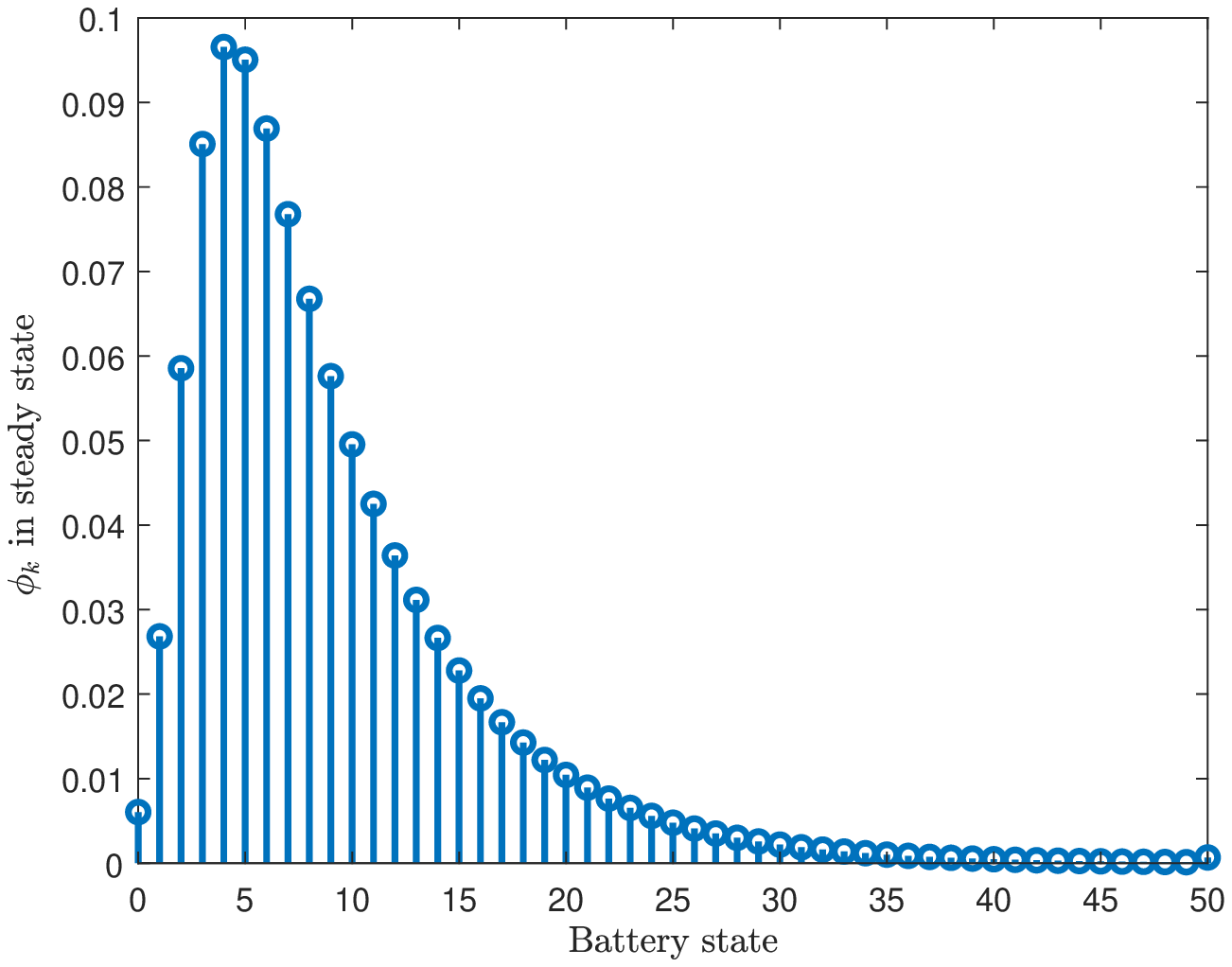}  
  \caption{$\rho=2, c_1 = 0.8$}
 \label{f5c}
\end{subfigure}
\begin{subfigure}[t]{0.24\textwidth}
  \centering
  \includegraphics[scale=.3]{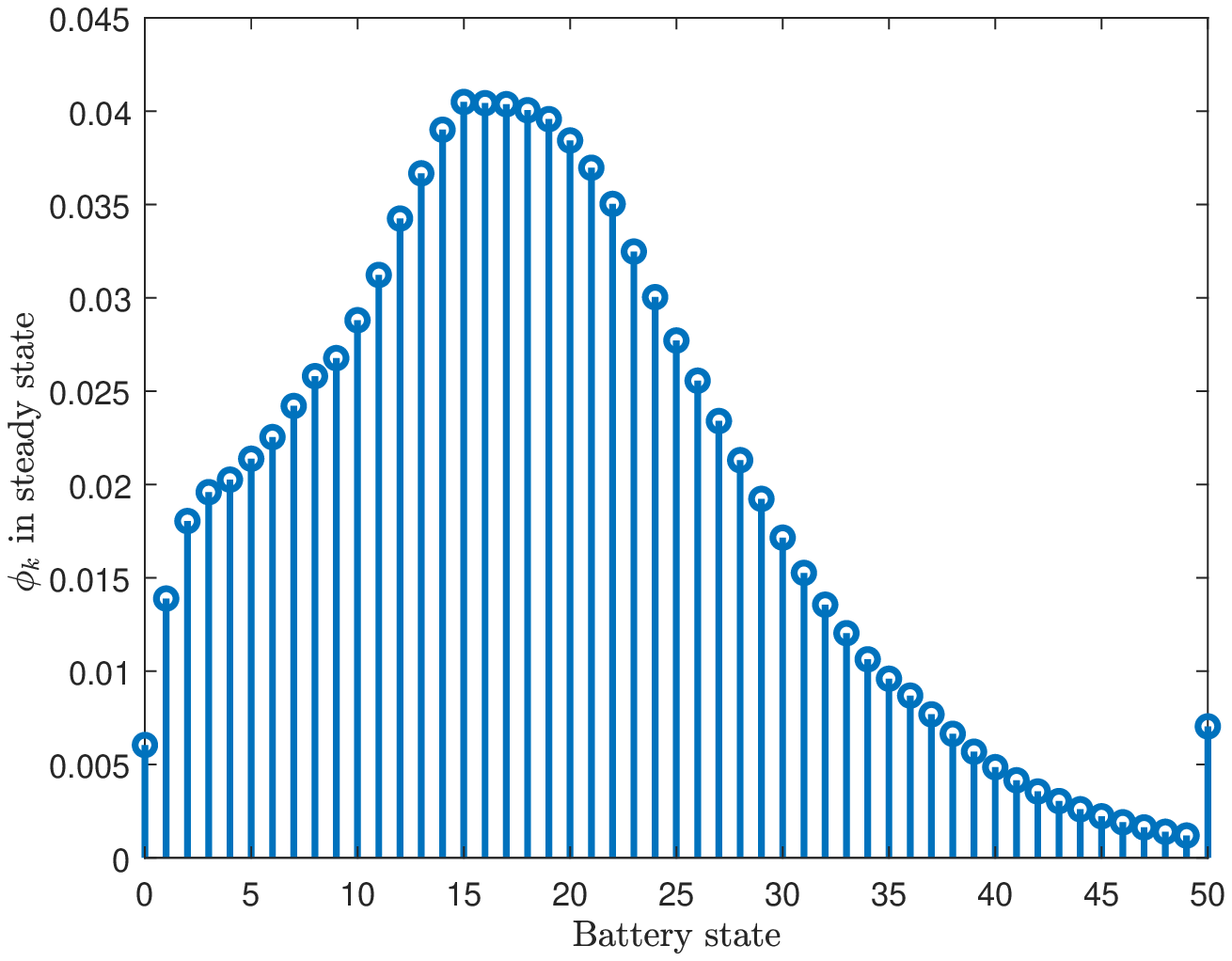}  
  \caption{$\rho=2, c_1 = 0.3$}
  \label{f5d}
\end{subfigure}
\caption{$K=50,\theta_1 = 3, \mu_{1,1} = 1, \gamma_{g_1}=2, c_2=1$}
\label{f5}
\end{figure}
\begin{table}[t]
\center
\scriptsize
\begin{tabular}{  |  m{5.9em}|m{4em} | m{4em} |m{4em} |m{4em} |  } 
\hline
\backslashbox[7.7em]{\textbf{\scriptsize  Probability}} {\textbf{\scriptsize  Cases}}& (a) $\rho\!=\!2$ $~~c_1\!=\!0.5$ &(b) $\rho\!=\!3$ $~~c_1\!=\!0.5$ & (c) $\rho\!=\!2$ $~~c_1\!=\!0.8$ &(d) $\rho\!=\!2$ $~~c_1\!=\!0.3$ \\
\hline
$\Pr (B_1\!=\!0)$ & 0.006 & 0.002&0.007&0.006 \\ 
$\Pr (B_1\!=\!50)$& 0.0015 & 0.017& 0.0006&0.007 \\ 
$\overline{B}_1$& 12.64 & 18.52& 8.84&19.37 \\ 
\hline
\end{tabular}
\caption{The values of $\Pr(B_1=0),~\Pr(B_1=50),  ~\overline{B}_1$ for cases (a), (b), (c), (d) in Fig.~(\ref{f5}).}
\label{table1}
\end{table}
\par $\bullet$ {\bf Behavior of probability vector $\boldsymbol{\Phi}$ in terms of $\boldsymbol{c_1}$ and $\boldsymbol{\rho}$}: In this part, we let $N=1$ assume that the FC applies a binary channel gain quantizer. We illustrate how the entries of vector $\boldsymbol{\Phi}$ in \eqref{inf} depend on the transmission coefficient $c_1$ in \eqref{alpha} and the energy harvesting parameter $\rho$. Fig.~(\ref{f5}) plots the entries of $\boldsymbol{\Phi}$ versus $k$ for $K=50,~ \gamma_{g}=2, ~c_2=1,~ \theta=3,~ \mu=1$. To accentuate the effect of $c_1$ in \eqref{alpha} and $\rho$ on the entries of $\boldsymbol{\Phi}$ we define the average energy stored at the battery of sensor $n$ as
\begin{equation}\label{ave_battery}
   \overline{B}_n=\mathbb{E}\{B_n\}=\sum_{k=0}^K k~\phi_{n,k} 
\end{equation}
where the largest possible value for $\overline{B}_n$ is $K$. Table \ref{table1} tabulates $\Pr(B_1=0),~ \Pr(B_1=K), ~\overline{B}_1$ for four cases, $(a)~\rho=2,~ c_1=0.5,~ (b)~\rho=3,~ c_1=0.5,~ (c)~\rho=2,~ c_1=0.8, ~(d)~ \rho=2, ~c_1=0.3$. Going from $(a)$ to $(b)$,  we note that as $\rho$ increases $\overline{B}_1$ increases, which is expected. Going from $(a)$ to $(c)$, we notice that as $c_1$ increases, $\overline{B}_1$ decreases. 
This is because as $c_1$ increases, data transmit energy $\alpha^2$ in \eqref{alpha} increases. Due to large energy energy consumption for data transmission (compared to energy harvesting) $\overline{B}_1$ decreases and the sensor may stop functioning, due to energy outage. 
Going from $(a)$ to $(d)$, we note that as $c_1$ decreases, $\overline{B}_1$ increases.
This is because as $c_1$ decreases, data transmit energy $\alpha^2$ in \eqref{alpha} decreases. Due to small energy energy consumption for data transmission (compared to energy harvesting)  $\overline{B}_1$ increases, indicating that the sensor has excess stored energy.
\par
\begin{figure}[!t]
\centering
\includegraphics[scale=.45]{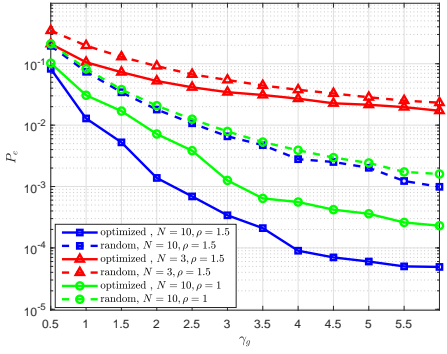}
 \caption{$P_e$ versus $\gamma_{g_n}$ for $K = 3,~ c_1 = 0.5,~ c_2=1,~ SNR_s=2.5$ dB.}
\label{f6}
\end{figure}

 $\bullet$ {\bf Solving $\boldsymbol{(\mathcal{P}_1)}$ and effectiveness of the optimization}: In this part, we focus on solving $(\mathcal{P}_1)$ and we demonstrate the effectiveness of optimizing the parameters $\{\theta_n, \mu_{n,l}\}, \forall n,l$ on enhancing the system error probability, when the FC applies a binary channel gain quantizer. In particular, we compare $P_e$ at the FC (the error probability in \eqref{pe1} corresponding to the optimal Bayesian fusion rule), when the optimization variables $\{\theta_n, \mu_{n,1}\}, \forall n$ are obtained from solving $(\mathcal{P}_1)$, denoted as ``optimized", and when these variables are some randomly chosen values, denoted as ``random" in the figures. 
We assume that the statistics of fading coefficients, the communication channel noise variances, and the observation noise variances for all sensors are identical in \eqref{y_n,t}, and define $SNR_s=20\log(\mathcal{A}/\sigma_v)$, as the SNR corresponding to observation channel. Fig.~(\ref{f6}) plots $P_e$ versus $\gamma_{g}$ for $\sigma_{w}^2$= 1, $\alpha_0=2$ (where $\alpha_0$ is the average transmit power per sensor constraint in $(\mathcal{P}_1)$), $N=3,~10$ and $\rho=1,~1.5, ~SNR_s=2.5$ dB. 
\par
\begin{figure}[!t]
\centering
\includegraphics[scale=.45]{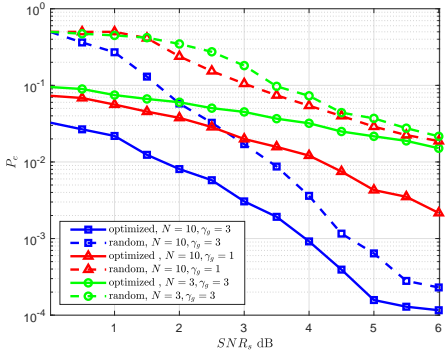}
 \caption{$P_e$ versus $SNR_s$ dB for $K = 3,~ \rho=2,~ c_1 = 0.5, ~c_2=1$.}
\label{f7}
\end{figure}

 For the ``optimized"  curve, we first obtain the optimization variables $\{\theta_n, \mu_{n,1}\}, \forall n$ from solving $(\mathcal{P}_1)$, and then use these optimized values to implement the system and run the Monte Carlo simulation.
For the ``random"  curve, we let $\theta_n=3,~\mu_{n,1}=1, \forall n$ for implementing and running the Monte Carlo simulation. 
To calculate each point on the ``optimized'' curve, we have conducted 10000 Monte-Carlo trials. In each trial, we generate fading channel and noise realizations, find the corresponding received vector, form the optimal Bayesian fusion rule, and make a decision. We let $P_e$ be the ratio of number of decision errors in 10000 Monte-Carlo trials. 
 Fig.~(\ref{f6}) shows that, given $N$, as $\rho$ increases, the gap between ``optimized" and ``random'' curves increases. Also, given $\rho$, as $N$ increases, this gap increases. 
 Fig.~(\ref{f7}) shows $P_e$ versus  $SNR_s$ for $\sigma_w^2=1,~ \alpha_0=2,~ N=3,~10, ~\rho=2, \gamma_g=1,~3$. Given $N=10$, as $\gamma_g$ increases, the gap between ``optimized" and ``random'' curves decreases. On the other hand, given $\gamma_g=3$, as $N$ increases, the gap between ``optimized" and ``random'' curves increases. 
\par
\begin{figure}[!t]
\centering
\includegraphics[scale=.45]{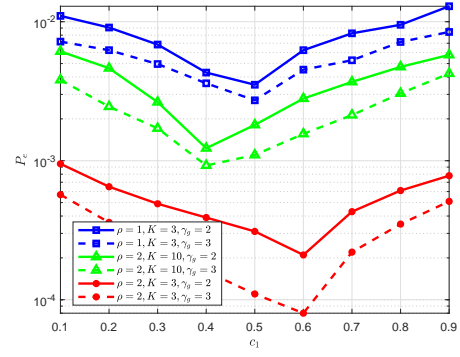}
 \caption{$P_e$ versus $c_1$ for $N=10, ~c_2=1,~  SNR_s=2.5$ dB.}
\label{f8}
\end{figure}

\begin{figure}[!t]
\centering
\includegraphics[scale=.45]{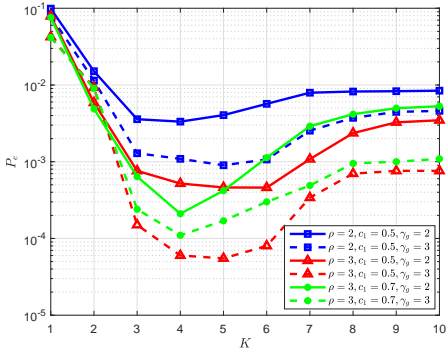}
 \caption{$P_e$ versus $K$ for $N=10,~c_2=1, ~\alpha_0=2,~SNR_s=2.5$dB.}
\label{f9}
\end{figure}

\begin{figure}[!t]
\centering
\includegraphics[scale=.45]{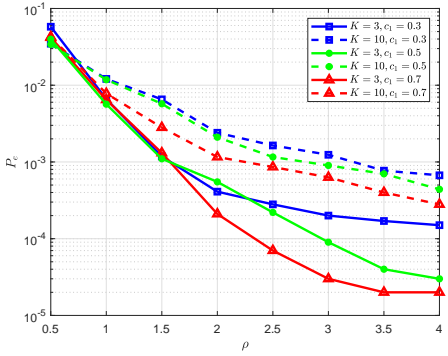}
 \caption{$P_e$ versus $\rho$ for $N=10,~ \rho=2, ~c_1 = 0.5, ~c_2=1, SNR_s=2.5$ dB.}
\label{f10}
\end{figure}

 $\bullet$ {\bf Behavior of $\boldsymbol{P_e}$ of the optimized system in terms of different parameters}:
Fig.~(\ref{f8}), Fig.~(\ref{f9}) and Fig.~(\ref{f10}) illustrate the behavior of $P_e$ of the optimized system as different parameters change. Fig.~(\ref{f8}) shows $P_e$ versus $c_1$ in \eqref{alpha}. In this simulation we assume $\rho=1,~2, ~K=3,~10, ~\gamma_g=2,~3,~ c_2=1.$
This figure suggests that, given $\rho,~ K,~\gamma_g$, there is an optimal $c_1$ value, which we denote as  $c_1^*$, that minimizes $P_e$. Starting from small values of $c_1$, as  $c_1$ increases (until it reaches the value $c_1^*$), $P_e$  decreases. 
This is because the harvested energy can recharge the battery and can compensate for the increase of data transmission power $\alpha^2$ in \eqref{alpha}.
However, when $c_1$ exceeds $c_1^*$, the harvested and stored energy can no longer support the increase of  data transmission power $\alpha^2$, and hence $P_e$ increases. 
We note that the value  $c_1^*$ depends on the values of $\rho,~K,~\gamma_g$. For instance, for $\rho=1,~K=3, \gamma_g=2,~3$, we have $c_1^*=0.5$.
Given $\rho$ and $\gamma_g$, as $K$ increases (green curves), $c_1^*$ decreases. On the other hand, given $\gamma_g$ and $K$, as $\rho$ increases (red curves), $c_1^*$ increases.
\par Fig.~(\ref{f9}) shows $P_e$ versus $K$ for $N=10,~c_2=1,~SNR_s =2.5$ dB, $\alpha_0=2$ and $c_1=0.5,~ 0.7, ~ \rho=2,~3,~ \gamma_g=2,~3.$
This figure indicates that, given $\rho,~c_1,~ \gamma_g$, there is an optimal $K$ value, which we denote as  $K^*$, that minimizes $P_e$. Starting from small values of $K$, as $K$ increases (until it reaches the value $K^*$), $P_e$  decreases. 
However, when $K$ exceeds $K^*$, $P_e$ increases, until it reaches an error floor. This is because for small $K$, data transmission power $\alpha^2$ is constrained by $K$ (small $K$ value leads to small $b_{n,t}$ values). Hence, as $K$ increases, $b_{n,t}$ values and thus $ \alpha^2$ increase, which cause $P_e$ to decrease. As $K$ exceeds $K^*$,
data transmission power $\alpha^2$ is no longer restricted by $K$, and instead it is restricted by $\rho$. Since the harvested energy is insufficient (the average  energy stored at the battery defined in \eqref{ave_battery} is small), $P_e$ increases. This trend continues, until the effect of communication channel noise dominates the detection performance and causes an error floor. 
We note that the value  $K^*$ depends on the values of $\rho,~ c_1, ~\gamma_g$. For instance, for $\rho=3,~ c_1=0.5, ~\gamma_g=3$, we have $K^* = 5$.
\par Fig.~(\ref{f10}) shows $P_e$ versus $\rho$ for $K=3,~10, ~N=10, ~\rho=2, ~c_1=~0.3,~ 0.5,~ 0.7$. We note that, given $K$ and $c_1$, as $\rho$ increases, $P_e$ decreases, until it reaches an error floor. Increasing $\rho$ any further, does not lower $P_e$. This is because for small $\rho$, data transmission power $\alpha^2$ is restricted by the amount of harvested energy. Hence, increasing $\rho$ decreases $P_e$. On the other hand, for large $\rho$, where $\Pr(B_n=K)$ is large, data transmission power $\alpha^2$ is not limited by energy harvesting. Instead the detection performance is limited by the communication channel noise, such that the smaller the communication channel noise variance is, the lower the error floor becomes.
\par
\begin{figure}[!t]
\centering
\includegraphics[scale=.45]{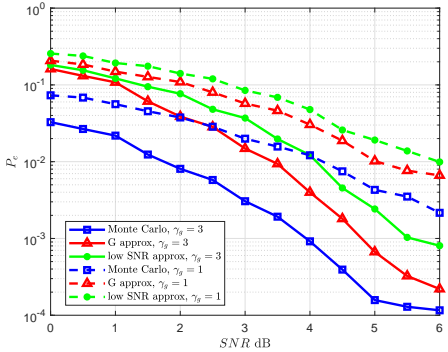}
 \caption{$P_e$ versus $SNR$ dB for $K = 3, N=10, ~\rho=2,~c_1 = 0.5,~ c_2=1,~ \alpha_0=2$.}
\label{f11}
\end{figure}

\begin{figure}[!t]
\centering
\includegraphics[scale=.45]{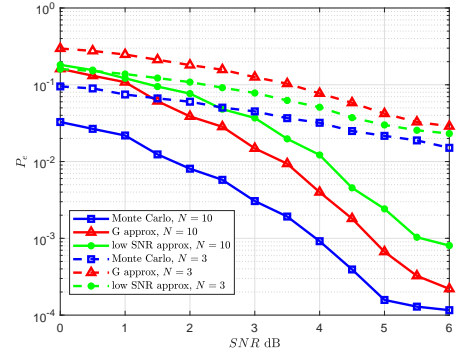}
 \caption{$P_e$ versus $SNR$ dB for $K = 3, ~\rho=2,~ \gamma_{g}=3,~c_1 = 0.5,~ c_2=1,~ \alpha_0=2$.}
\label{f12}
\end{figure}

 $\bullet$ {\bf Accuracy of two error probability approximations in \eqref{snr_pe} and \eqref{clt_pe}}:
In this part, we examine the accuracy of the two $P_e$ approximations we provided in \eqref{snr_pe} and \eqref{clt_pe}, denoted as ``low SNR approx." and ``G approx.", compared to exact $P_e$ obtained from Monte Carlo simulations, denoted as ``Monte Carlo" in the figures. We define $SNR=-20\log(\sigma_w)$.
Fig.~(\ref{f11}) plots $P_e$ versus SNR for $SNR_s=2.5$ dB, $K=3,~N=10,~\rho=2,~\gamma_g=1,~3,~\alpha_0=2,~c_1=0.5$ and $c_2=1$. This figure shows that given $\gamma_g$, as $SNR$ increases, $P_e$ , ``G approx.''  and ``low SNR approx.'' decrease. Also, ``G approx." is a better approximation than ``low SNR approx.". 
Fig.~(\ref{f12}) plots $P_e$ versus SNR for $SNR_s=2.5$ dB, $K=3,~N=3,~10,~\rho=2,~\gamma_g=3,~\alpha_0=2,~c_1=0.5$ and $c_2=1$. This figure indicates that given $N$, as $SNR$ increases, $P_e$ , ``G approx.'' and ``low SNR approx.'' decrease. While for $N=10$ ``G approx." is a better approximation than ``low SNR approx.", for $N=3$, ``low SNR approx." is a better approximation than ``G approx.".
\par
\begin{figure}[!t]
\centering
\includegraphics[scale=.45]{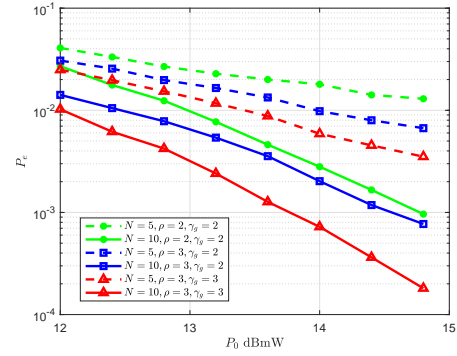}
 \caption{$P_e$ versus $P_0$dBmW for $K = 3,~ c_1 = 0.5,~ c_2=1$.}
\label{f13}
\end{figure}

 $\bullet$ {\bf Effect of random deployment of sensors on $\boldsymbol{P_e}$ of the optimized system}: 
To investigate the effect of random deployment on $P_e$ of the optimized system, we consider $(\mathcal{P}_1)$, with the difference that, for $J_n$  in \eqref{sim_j}, $P_{\text{d}_n}$ and $P_{\text{f}_n}$ expressions are replaced with the ones in \eqref{pd_pf2}. We assume that sensors are randomly deployed in a circle field with radius $r_1=100$ meters, the signal source with power $P_0$ is located at the center of this field, and it is at least $r_0=1$  meter away from any sensor within the field. Fig.~(\ref{f13}) illustrates $P_e$ versus $P_0$ for $\alpha_0=2,~ \gamma_g=2,~3,~ \rho=2,~3$. We note that as $P_0$ increases, $P_e$ decreases. Also, given $\rho$ and $\gamma_g$, the rate of decrease in $P_e$ increases as $N$ increases.   
\par
\begin{figure}[!t]
\centering
\includegraphics[scale=.45]{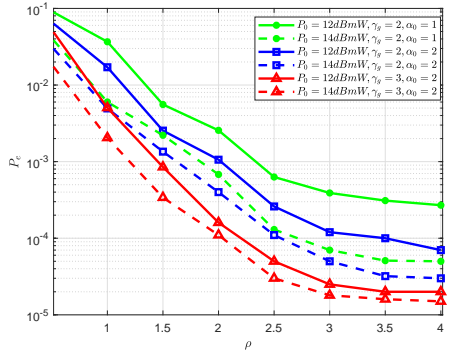}
 \caption{$P_e$ versus $\rho$ for $K = 3,~ N=10,~ c_1 = 0.5, ~c_2=1$.}
\label{f14}
\end{figure}

 Fig.~(\ref{f14}) demonstrates  the behavior of $P_e$  of the optimized system as different parameters change. This figure shows $P_e$ versus $\rho$ for $P_0=12,~14$ dBmW, $\gamma_{g}=2,~3$, $\alpha_0=1,~2,~K=3,~c_1=0.5,~c_2=1$. 
We notice that,  given $P_0$, $\gamma_g$ and $\alpha_0$, as $\rho$ increases, $P_e$ reduces, until it reaches an error floor. Increasing $\rho$ any further, does not lower $P_e$. This behavior is similar to Fig.~(\ref{f10}) and it is because for small $\rho$,  data transmission power $\alpha^2$ is restricted by the amount of harvested energy. Hence, increasing $\rho$ reduces $P_e$. On the other hand, for large $\rho$, data transmission power $\alpha^2$ is not limited by energy harvesting any more and it is limited by the communication channel noise.
\par $\bullet$ {\bf Solving $\boldsymbol{(\mathcal{P}_2)}$ and effectiveness of the optimization}: In this part, we focus on solving $(\mathcal{P}_2)$ and we illustrate the effectiveness of optimizing the parameters $\{\theta_n, \mu_{n,l}\}, \forall n,l$ on reducing the average total transmit power, denoted as $P_{tot}$ here, when the FC applies a binary channel gain quantizer. In
particular, we compare  $P_{tot}$  when
the optimization variables $\{\theta_n, \mu_{n,l}\}, \forall n,l$  are obtained from
solving $(\mathcal{P}_2)$, denoted as ``optimized”, and when these variables are some randomly chosen values, denoted as ``random” in the figures. We assume that the statistics of fading coefficients, the communication channel noise variances, and the observation noise variances for all sensors are identical in \eqref{y_n,t}.
\par
\begin{figure}[!t]
\centering
\includegraphics[scale=.45]{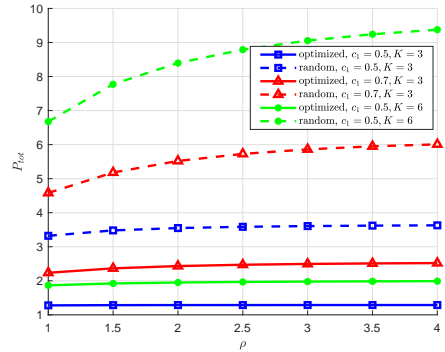}
 \caption{$P_{tot}$ versus $\rho$ for $N=10, ~c_2=1, ~J_0=1.2$.}
\label{f15}
\end{figure}

 Fig.~(\ref{f15}) shows $P_{tot}$ versus $\rho$ for $N=10,~K=3,~6,~c_1=0.5,~0.7,~ c_2=1,~J_0=1.2,~\gamma_g=2,~\sigma_{w}^2=1,~ SNR_s=2.5$ dB. For the ``random"  curve, we let $\theta_n=3, \mu_{n,1}=1, \forall n$. 
We note that given $K$ and $c_1$, as $\rho$ increases, $P_{tot}$ increases until it reaches a ceiling, and the ceiling value depends on $K$ and $c_1$. Increasing $\rho$ any further, does not increase $P_{tot}$. This is because for small $\rho$, data transmission power $\alpha^2$
is restricted by the amount of harvested energy. Hence, increasing $\rho$ increases $P_{tot}$. On the other hand, for large $\rho$, data
transmission power $\alpha^2$ is not limited by energy harvesting. Instead it is limited by the battery size $K$ and transmission coefficient $c_1$. 
This figure also shows that given $c_1$, as $K$ increases, the gap between ``optimized" and ``random" curves increases. Also, given $K$, as $c_1$ increases, the gap between ``optimized" and ``random" curves increases.
\par
\begin{figure}[!t]
\centering
\includegraphics[scale=.45]{ 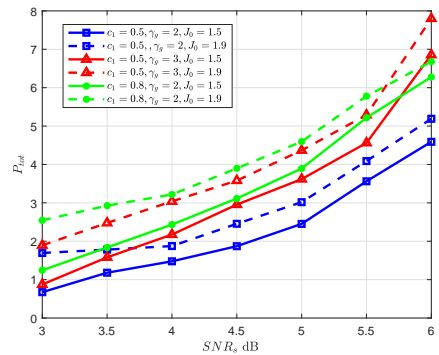}
 \caption{$P_{tot}$ versus $SNR_s$ dB for $K = 3, N=10, \rho=1, c_2=1$.}
\label{f16}
\end{figure}

 $\bullet$ {\bf Behavior of the average total transmit power of the optimized system in terms of different parameters}:
Fig.~(\ref{f16}) shows $P_{tot}$ versus $SNR_s$ for $K=3,~ N=10,~\rho=1,~\gamma_g=2,~3,~ J_0=1.5,~1.9,~c_1=0.5,~0.8,~c_2=1$.
 This figure shows that, given $\gamma_g$ and $J_0$, as $c_1$ increases, data transmission power $\alpha^2$ in \eqref{alpha} and therefore, $P_{tot}$ increase. Given $\gamma_g$ and $c_1$, as $J_0$
 in the constraint of $(\mathcal{P}_2)$ increases, $P_{tot}$ increases. Recall that the $J$-divergence and $P_e$ are related through $P_e>\Pi_0 \Pi_1 e^{-J/2}$ \cite{vin}. Hence, increasing $J_0$ implies that $(\mathcal{P}_2)$ should be solved subject to a smaller $P_e$ value constraint. To satisfy a tighter constraint on $P_e$, $\alpha^2$ in \eqref{alpha} and hence $P_{tot}$ increase. 
 Given $J_0$ and $c_1$, as $\gamma_g$ increases, $\alpha^2$ in \eqref{alpha} and hence, $P_{tot}$ increase. This is because as $\gamma_g$ increases, $\Pr( g^2 \in (\mu_1^2, \infty))=e^{-\mu_1^2/\gamma_g}$ increases as well. Considering that $c_2> c_1$, this probability increase leads to an increase in $\alpha^2$ and thus, an increase in $P_{tot}$.
\section{Conclusions}\label{conclu}
In this work, we studied design and performance of a WSN with EH-powered sensors and the FC, tasked with detecting a known signal in uncorrelated Gaussian noises. 
We proposed a power adaptation scheme, which allows each sensor to intelligently choose its transmission symbol, such that the larger its stored energy and its quantized channel gain are, the higher its transmit power is. Modeling the randomly arriving energy units during a time slot as a Poisson process and the dynamics of the battery as a finite state Markov chain, we formulated two problems $(\mathcal{P}_1)$ and $(\mathcal{P}_2)$, where the optimization parameters (the local decision thresholds and the channel gain quantization thresholds)
are embedded in the proposed power adaptation scheme, and play key roles in balancing energy harvesting and  energy consumption in data transmission. 
Our numerical results demonstrated the effectiveness of our optimization on enhancing the detection performance of the optimal Bayesian fusion rule in $(\mathcal{P}_1)$, and lowering the average total transmit power in $(\mathcal{P}_2)$. They also illustrated how the combination of energy harvesting rate $\rho$, the battery size $K$, the sensor observation and communication channel parameters ($SNR_s, ~SNR, ~\gamma_g$) impact our obtained solutions and the system performance.
In particular,  we observed that,  given $\rho$ and the observation and channel parameters, there is an optimal $K$ that minimizes the detection error. 
We also showed that, increasing $\rho$ does not always reduce the detection error. Depending on the strength of the communication channel noise, the detection error can reach an error floor, even for large $K$ and $\rho$.  
We also examined how the random deployment of sensors affects the optimization solutions and the system performance. 
\bibliographystyle{IEEEtran}
\bibliography{RefEH}
\end{document}